\documentclass[]{interact}
\usepackage{epstopdf}
\usepackage{subfigure}
\usepackage{lineno}
\usepackage{color,soul}

\usepackage{natbib}
\bibpunct[, ]{(}{)}{;}{a}{}{,}
\renewcommand\bibfont{\fontsize{10}{12}\selectfont}

\theoremstyle{plain}

\theoremstyle{definition}

\theoremstyle{remark}

\begin{document}

\articletype{ARTICLE TEMPLATE}

\title{A flock-like two-dimensional cooperative vehicle formation model based on potential functions}

\author{Ruochen Hao$^{\rm a,b}$, Meiqi Liu$^{\rm a}$, Wanjing Ma$^{\rm b}$, Bart van Arem$^{\rm a}$, Meng Wang$^{\rm a}$$^{\ast}$ \thanks{$^\ast$Contact Meng Wang. Email m.wang@tudelft.nl	\vspace{6pt}}
	\\\vspace{6pt}  $^{a}${\em{Department of Transport and Planning, Delft Univeristy of Technology, The Netherlands}}; \\
	$^{b}${\em{Key Laboratory of Road and Traﬃc Engineering of the Ministry of Education, Tongji University, 4800 CaoAn Road, Shanghai, China}}
}
\maketitle

\begin{abstract}
Platooning on highways with connected and automated vehicles (CAVs) has attracted considerable attention,  while how to mange and coordinate platoons in urban networks remains largely an open question. This scientific gap mainly results from the maneuver complexity on urban roads, making it  difficult to model the platoon formation process. Inspired by flocking behaviors in nature, this paper proposed a two-dimensional model to describe CAV group dynamics. The model is formulated based on potential fields in planar coordinates, which is composed of the inter-vehicle potential field and the cross-section potential field. The inter-vehicle potential field enables CAVs to attract each other when vehicle gaps are larger than the equilibrium distance, and repel each other otherwise. It also generates incentives for lane change maneuvers to join a platoon or to comply with the traffic management layer. The cross-section potential field is able to mimic lane keeping behavior and it also creates  resistance to avoid unnecessary lane changes at very low incentives. These modeling principles can also be applied to human-driven vehicles in the mixed traffic environment.  Behavioral plausibility of the model in terms of car-following rationality and car-following safety is demonstrated analytically and further verified with simulation in typical driving scenarios. The model is computationally efficient and can provide insights into platoon operations in urban networks.
\end{abstract}


\section{Introduction}
Connected and automated vehicles (CAVs) have the potential to affect the traffic flow characteristics and traffic management and control \citep{Pei2019}. Unlike individual automated vehicles which rely solely on on-board sensors, CAVs have advantages in both situation awareness and cooperative maneuvering \citep{wang2014rolling}. 
As one of the earliest CAV applications, platooning has attracted a lot of attention due to the underlying benefits of improving roadway capacity and energy economy. Earlier studies in Intelligent Vehicle Highway System (IVHS)  demonstrated  that  highway capacity can be improved by applying platooning \citep{Tan1998}.

The architecture of IVHS is composed of the network layer, link (traffic control) layer, the platoon planning/management layer,  the platoon/vehicle control layer, and the physical layer \citep{horowitz2000control,baskar2011traffic}. Platoon management layer deals with maneuver strategies of platoon formation, merge, split, dissolution, in addition to platoon member selection \citep{Jia2015,Shladover2016Cooperative}. On the contrary, the platoon control layer focuses on motion control, which transfers management strategies to executable trajectories. Existing studies focused on platoon control layer, resulting in significant contributions on string-stable platooning controller design \citep{Milanes2014,Shladover2016Cooperative}, with advances in dealing with delays \citep{zhang2020control,jin2016optimal}, uncertainties \citep{chen2018robust}, and the heterogeneous platoon \citep{wang2014rolling}. For extensive review of the control architectures, the control methods and string stability analysis of the platoon control layer, we refer to \citet{wang2018review}.

Modeling and control design at the platoon management layer receives much less attention compared with the platoon control layer. Active route planning strategies for platoon formation  focused on optimizing routing and departure time to increase the chance of CAVs meeting each other along their routes \citep{Turri2016,van2016fuel}. But the operational processes to form the platoon are not considered. Although platoon operations can be interrupted by intersections on urban roads, forming platoons can still be favorable since it not only improves the throughput of the road network but also reduces the complexity of the traffic control layer, i.e.\ the platoon of several vehicles can be treated as one control unit. There are a number of studies optimizing platoon trajectories at signalized intersections assuming  that vehicles already form platoons when entering the network \citep{han2020energy,chen2015platoon}. The research summarized in \citet{rios2016survey} excludes the platoon formation under the urban environment. 

Limited studies consider active platoon management strategies at signalized intersections. Existing research efforts on platoon formation are rule-based, for instance, according to the signal timing \citep{Faraj2017} or the communication range \citep{Jin2013}. \citet{Tallapragada2017} uses K-mean clustering to form platoons in order to ensure the size of formed platoons is similar to the requirement of the intersection traffic controller. Decentralized control is proved to be economical \citep{otto2009distributed}. The  decentralized contorl reported by \citet{halle2004decentralized} focuses more on the process of merging and splitting, and did not consider traffic signal control. Moreover, the mixed traffic environment is ignored. \citet{yao2020managing} proposes a decentralized platoon strategy based on the maximal platoon size on urban roads under the mixed traffic environment. However, the lateral lane change behavior is not considered.   
 
Swarm intelligence is an important decentralized control method which has been widely used in unmanned aerial vehicles (UAVs) related research. Swarm intelligence control is a control method that allows the overall group to reach the desired target by setting local interaction rules for each agent according to several principles. Swarm intelligence can be used for both platoon formation and tracking \citep{ren2010distributed}. With respect to formation, the matrix-based approach and the Lyapunov-based approach are included \citep{cao2012overview}. As for formation tracking, the behavior-based approach and the potential field approach have been used \citep{das2016cooperative}. Swarm intelligence models originate from  the basic flock principles proposed by \citet{10.1145/37401.37406} in 1987. Existing flock-like models for UAVs mainly focus on selecting the platoon leader and building communication network \citep{Cooper2016}, while neglecting the potential field and detailed motions. Potential fields were applied in recent studies to control multi-vehicles motion \citet{yi2020using}. \citet{li2017extended} applied the spring-mass system theory to establish the car-following model based on the elastic potential field, and \citet{li2019new} considered time delay and stability analysis in the spring-mass-damper-clutch system. \citet{bang2017platooning} used swarm intelligence as the control strategy and spring-mass-damper system theory as agent interaction rules. But these are limited to longitudinal vehicle motion. Since the dynamic characteristics of on-road vehicles are different from the UAVs or other common research objects of swarm intelligent, existing potential field models of swarm intelligence can not be used directly to two-dimensional cooperative vehicle group dynamics. 

To summarize, although numerous efforts on platooning have been made, there are still some scientific gaps on platoon management and control at urban networks. Platoon formation is usually considered as a centralized macroscopic level problem \citep{Luo2018}, which plans vehicle routes and schedules to form platoons but ignores the detailed formation processes. Existing microscopic approaches do not fully utilize the flock-like model, thus vehicles are operated in one dimension with myopic horizon. As a result, limited studies consider the vehicle sequence adjustment \citep{Tallapragada2017}, especially in the urban area. Owing to the different turnings and routes of CAVs, the orders of CAVs are critical in the intersection control (i.e. adaption of signals based on the class-specific traffic demand levels can achieve better traffic  control performance). Additionally, most platoon formation methods only work in fully CAV environment and the influence of human-driven vehicles (HVs) is ignored in many studies. 

This paper proposed a flock-like vehicular behavior model with flexible formation in two dimensions. This model integrates the tactical maneuver into operational motion dynamics using the combined potential fields. The lane changing and car following behaviours are considered and the endogenous flocking principles of attraction (or flock centering), alignment (or velocity matching) and collision avoidance are captured in this model. Vehicles can form platoons according to  the requirements from the traffic control layer or based on vehicle  classes using the correlation coefficient in the potential fields. Furthermore, this model can represent both CAVs and  human-driven vehicles.

The remainder of this paper is organized as follows.  Section \ref{section:protocal}  describes the problem and presents the platoon management model. The design of potential field is located at this section. Section \ref{section:calibration} analyzes the model mathematical properties.  Section \ref{section:evolution} uses some numerical studies to evaluate the performance of the proposed flock-like model and explore the reasonable ranges of parameters. Finally, conclusions and recommendations are provided in Section \ref{section:conclusion}.

\section{Model formulation} \label{section:protocal}
\subsection{Problem description}
\begin{figure}
  \centering
  \resizebox*{8cm}{!}{\includegraphics{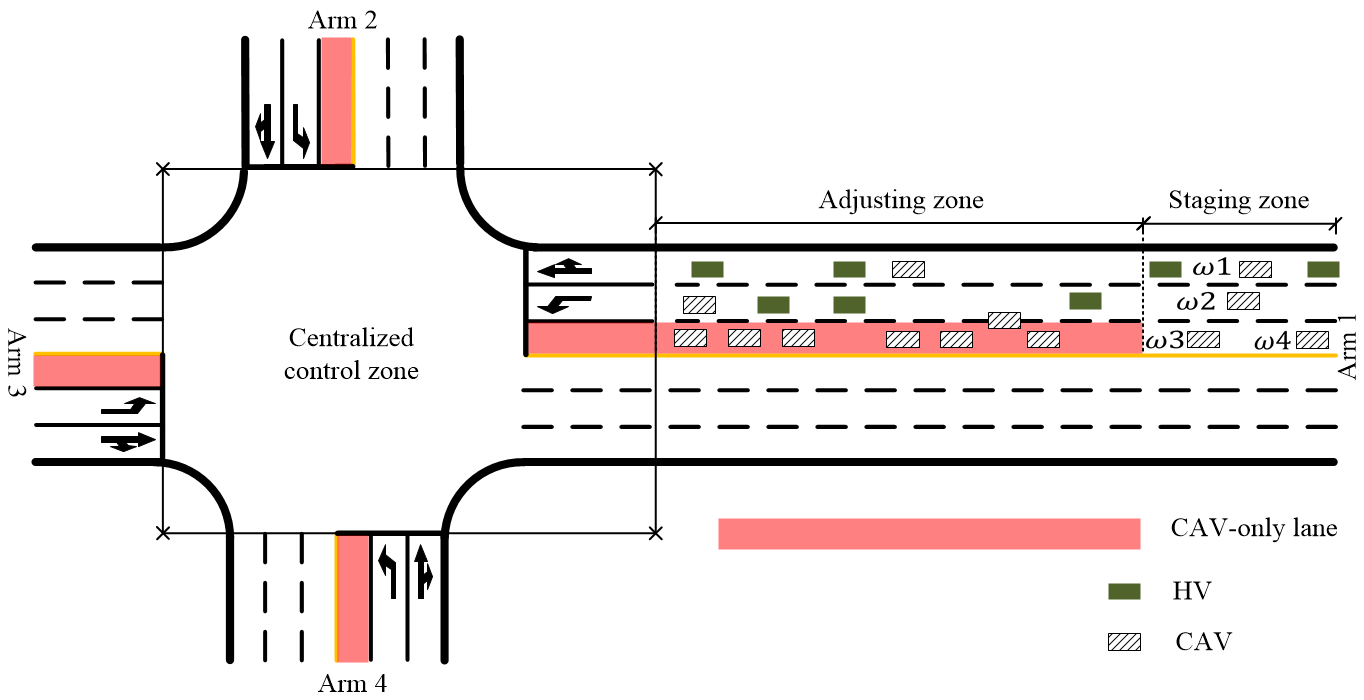}}
  \caption{Intersection description}\label{fig:Intersection-description}
\end{figure}

\begin{figure}
  \centering
  \resizebox*{8cm}{!}{\includegraphics{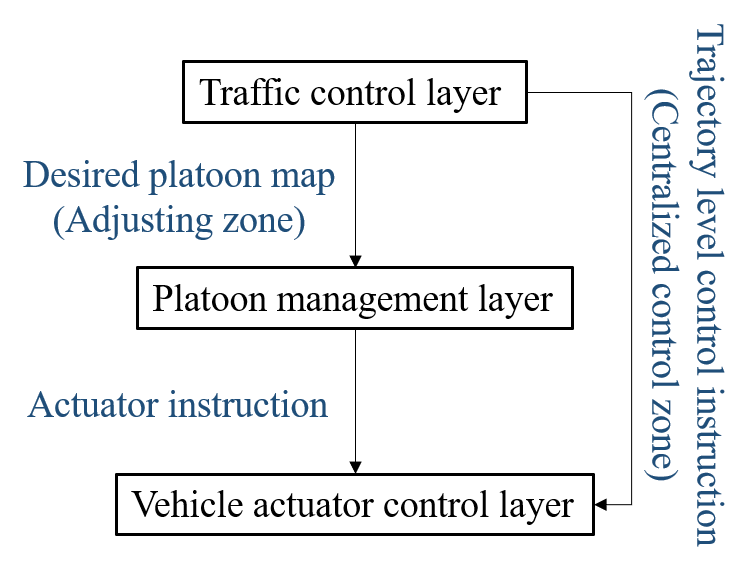}}
  \caption{Intersection control structure}\label{fig:Intersection-control-structure}
\end{figure}

The purpose of the proposed model is to describe the operation process of the formation of CAV platoons under the mixed traffic environment.  Figure~\ref{fig:Intersection-description} illustrates the context of this study, which is a full intersection with a CAV exclusive lane, a left-turn lane and a through/right-turn lane in each approach. The intersection is divided into three zones, i.e. the staging zone, the adjusting zone and the centralized control zone. The centralized traffic controller collects CAV information in the staging zone and provides CAVs with the desired platoon formation information such as the platoon size and the CAV sequence. CAVs in the adjusting zone will change their positions and velocities to track the commands of desired formation from the traffic control layer. After platoons enter the centralized control zone, the centralized traffic controller will give the control instructions to the CAVs directly. The control structure is shown as Figure~\ref{fig:Intersection-control-structure}. 

The focus of this paper is the platoon management layer in the adjusting zone. The proposed model describes the dynamic motions of both longitudinal and lateral dimensions to form the desired platoon formation in the adjusting zone, which is located at the platoon management layer.
Shown as an example in Figure~\ref{fig:Intersection-description}, four CAVs (i.e. $\omega_{1}$, $\omega_{2}$, $\omega_{3}$ and $\omega_{4}$) are in the staging zone. The centralized traffic controller requires $\omega_{2}$, $\omega_{3}$ and $\omega_{4}$ to form a platoon on the CAV exclusive lane, and consequently these CAVs adjust the longitudinal and lateral motions in the adjusting zone for this common purpose. 

\subsection{Modeling principles based on flocking}
The basic flocking model is proposed in \cite{10.1145/37401.37406}. The model is composed of three  principles: collision avoidance, velocity matching and flock centering. Collision avoidance guarantees collision-free with other flockmates, and flock centering ensures that each flock member keeps close to the neighboring flockmates and approaches to the perceived center. Velocity matching aims to keep the same velocity of the neighboring flockmates. 

Analogously, collision avoidance is one of the necessary requirements for CAVs. The velocity matching and flock centering behaviors are similar to the platoon stability in longitudinal direction as  CAVs in the same group/formation aim to keep the same velocity and space headway. Therefore, the flock-like model is suitable to describe the platoon operations. In the remainder of this section, we will introduce the vehicle dynamics and constraints, as well as the mathematical specifications of these flocking principles. 

\subsection{Vehicle dynamics and constraints}
The two-dimensional vehicle motion follows the basic law of kinematics at a point mass. Let vector $\mathbf{x}_j$ denote the system state of vehicle $j$, which consists of the longitudinal  and lateral positions ($x_j$ and $y_j$), as well as the longitudinal and lateral velocities ($v_{x,j}$ and $v_{y,j}$).
\begin{equation}
    \mathbf{x}_j = (x_j, v_{x,j}, y_j, v_{y,j})^T
\end{equation}
where $x$ is the longitudinal position in the driving direction along the road, and $y$ is the lateral position on the direction perpendicular to the road. The state dynamics follows the second-order kinematic equation as:
  \begin{equation}
  \label{eq:system_dynamics}
  \left(\begin{array}{c}
  \dot{x}_j\\
  \dot{v}_{x,j}\\
  \dot{y_j}\\
  \dot{v}_{y,j}
  \end{array}
  \right)=\left(\begin{array}{c}
  v_{x,j}\\
  a_{x,j}\\
  v_{y,j}\\
  a_{y,j}
  \end{array}
  \right)
  \end{equation}
where the longitudinal and lateral accelerations ($a_{x,j}$ and $a_{y,j}$) are model inputs. 
$v_x$, $a_x$ and $v_y$, $a_y$ are the velocity and acceleration on the longitudinal direction and the lateral direction respectively.

In the two-dimensional vehicle kinematic model, acceleration and velocity are constrained to respect the realistic vehicle dynamics, including: 

\begin{subequations}\label{eq:motion_constraints}
\begin{equation}
     a_x^{min} \leq  a_{x,j} \leq a_x^{max}
     \label{subeqnparta}
\end{equation}
\begin{equation}
     a_y^{min} \leq  a_{y,j} \leq a_y^{max}
     \label{subeqnpartb}
\end{equation}
\begin{equation}
     v_x^{min} \leq  v_{x,j} \leq v_x^{max}
     \label{subeqnpartc}
\end{equation}
\begin{equation}
     v_y^{min} \leq  v_{y,j} \leq v_y^{max} 
     \label{subeqnpartd}
\end{equation}
\end{subequations}

The velocity and acceleration limits are usually bounded within constant values. $a_x^{min}, a_y^{min}$ and $a_x^{max}, a_y^{max}$ are minimum and maximum longitudinal or lateral accelerations respectively. Similarly, $v_x^{min}, v_y^{min}$ and $v_x^{max}, v_y^{max}$ are minimum and maximum longitudinal or lateral velocities respectively. These parameters are determined by vehicle dynamics and road condition.

\subsection{Potential function specification}\label{potenrialfunction}
It is  assumed that CAVs perceive the states of other CAVs and HVs accurately in the perception range of their sensors and V2X devices, and the vehicle actuation systems in the lower level can precisely execute the control commands from the on-board controller.
The “mass” of vehicles in this study is set to the unit mass, which means the unit of force is $m/s^2$, i.e.\ the value of the lateral/longitudinal acceleration is equal to the value of the lateral/longitudinal force. Forces imposed on vehicles stem from interactions with surrounding vehicles and the deviation from lane center. These forces are considered as Potential Field Force (indicated by $F$). 

Let $\cal N$ and $N$ denote the set of vehicles in the perception range of the vehicle $j$, and the total number of vehicles in $\cal N$. The composite forces imposed on vehicle $j$ is:
  \begin{equation}
  F_j=\sum_{i=1, i\not=j}^{N} p_{x,ij}^b(x_i,v_{x,i},x_j,v_{x,j})+p_{y,ij}^b(y_i,y_j)+p_j^c(y_j)+p_j^d(v_{x,j})+ p_j^f
  \end{equation}

Here, $p_{x,ij}^b$ and $p_{y,ij}^b$ are the longitudinal and lateral inter-vehicle potential forces imposed on vehicle $j$ caused by vehicle $i$. The cross-section potential force, $p_j^c$, constrains the vehicle $j$ in the lateral dimension. Each CAV calculates $p_{x,ij}^b$ and $p_{y,ij}^b$ for all surrounding human-driven vehicles in the perception range of on-board sensors and other CAVs with the same platoon number within the communication range. CAVs with the same platoon number implies that the centralized traffic controller asks them to form a platoon, while the vehicle class can be used in the absence of a central traffic controller. Same vehicle class may imply vehicles have the same destination, target velocity or other features. HVs are assumed to estimate $p_{x,ij}^b$ and $p_{y,ij}^b$ considering all vehicles in the perception range of the driver.

Force $p_j^d$ is designed to operate the vehicle $j$ in the desired velocity.  If the value of $p_j^d$ is set as the maximal acceleration, the vehicle $j$ can accelerate to the desired velocity with the maximum acceleration providing no vehicles in front. In addition, the friction $p_j^f$ is also considered to avoid the local oscillation of vehicles in the lateral dimension, which is perpendicular to the driving direction. 

 The composite force imposed on the control object, $F_j$, is the sum of all forces, $p_{x,ij}^b$, $p_{y,ij}^b$, $p_j^c$, $p_j^d$ and $p_j^f$. $F_j$ is further decomposed into two perpendicular vectors, $F_{x}$ and $F_{y}$, as the model inputs $a_x$ and $a_y$  in order to update the motion of the subject vehicle. 

In the longitudinal direction, $p_j^d(v_{x,j})$ and $p_{x,ij}^b(x_i,v_{x,i},x_j,v_{x,j})$ act on the vehicle $j$, thus the longitudinal force/acceleration of vehicle $j$ is:
  \begin{equation}
  a_{x,j} = F_{j}^{x}= \sum_{i=1}^{N} p_{x,ij}^b(x_i,v_{x,i},x_j,v_{x,j})+p_j^d(v_{x,j})
  \end{equation}
where $p_j^d(v_{x,j})$ is set as:
 \begin{equation}
  p_j^d(v_{x,j})=\max\{F_{max}(\frac{v_x^{max}-v_{x,j}}{v_x^{max}}),0\}\label{eq:10-1}
  \end{equation}

In the lateral dimension, the lateral force/acceleration considering $p_{y,ij}^b(y_i,y_j)$, $p_j^c$, and $p_j^f$ is:
  \begin{equation}
  a_{y,j} = F_{j}^{y}= \sum_{i=1}^{N}p_{y,ij}^b(y_i,y_j)+p_j^c+p_j^f
  \end{equation}
  
It is noted that the longitudinal and lateral forces are used to update the vehicle motions under the system dynamics \eqref{eq:system_dynamics} and the motion constraints (\ref{eq:motion_constraints}). 

\subsubsection{Inter-vehicle potential function} 
\label{subsection:inter-vehicle}
Inter-vehicle potential function guarantees that vehicles in a group attract each other but maintain at least the safe gap. The longitudinal inter-vehicle potential function is specified as follows:
  \begin{equation}
  p_{x,ij}^b(x_i,v_{x,i},x_j,v_{x,j})=c_{ij}(\ln(x_{ij})-\frac{(x_{e}-t_{h}\Delta v_{x,ij})\ln(x_{e}-t_{h}\Delta v_{x,ij})}{x_{ij}})
  \label{eq:inter-vehicle-long}
  \end{equation}
where $x_{e}$ is the initial equilibrium distance in the longitudinal dimension. $x_{ij} = x_i - x_j$ is the longitudinal distance between vehicle $i$ and vehicle $j$. $\Delta v_{x,ij}=v_{x,i}-v_{x,j}$ is the relative longitudinal velocity. $c_{ij}$ is the correlation coefficient between the vehicle pair. For vehicles in the same group, the correlation coefficients between each vehicle pair are larger than the counterparts from different groups, leading to larger attractive forces to form a platoon. $t_h$ can be interpreted as  desired time to collision. $p_{x,ij}^b$ in the longitudinal direction is calculated by the aforementioned parameters and variables. 

\begin{figure}[h]
  \centering
  \resizebox*{10cm}{!}{\includegraphics{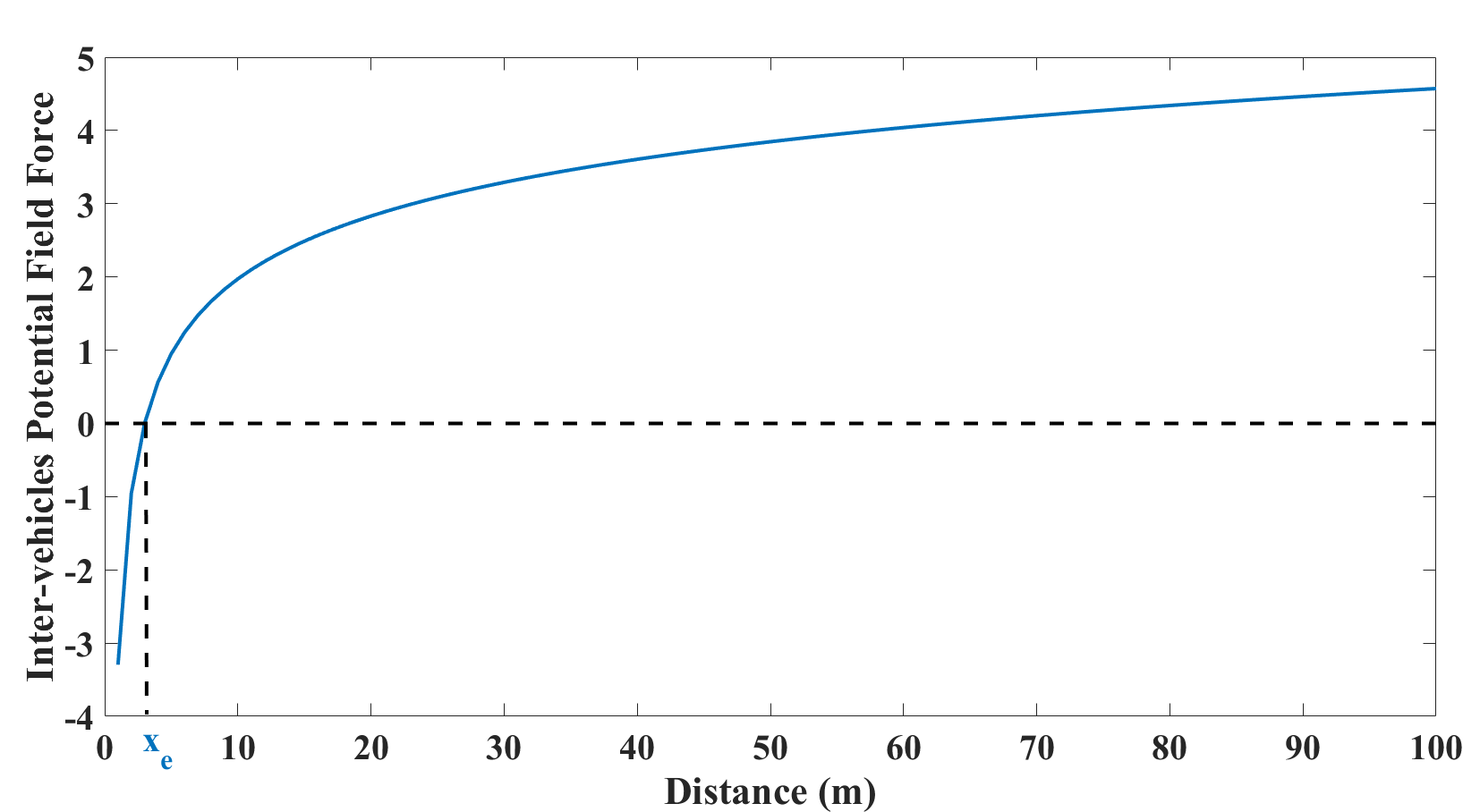}}
  \caption{Inter-vehicle potential function}\label{fig:Inter-vehicle-function}
\end{figure}

\begin{figure}[h]
  \centering
  \resizebox*{10cm}{!}{\includegraphics{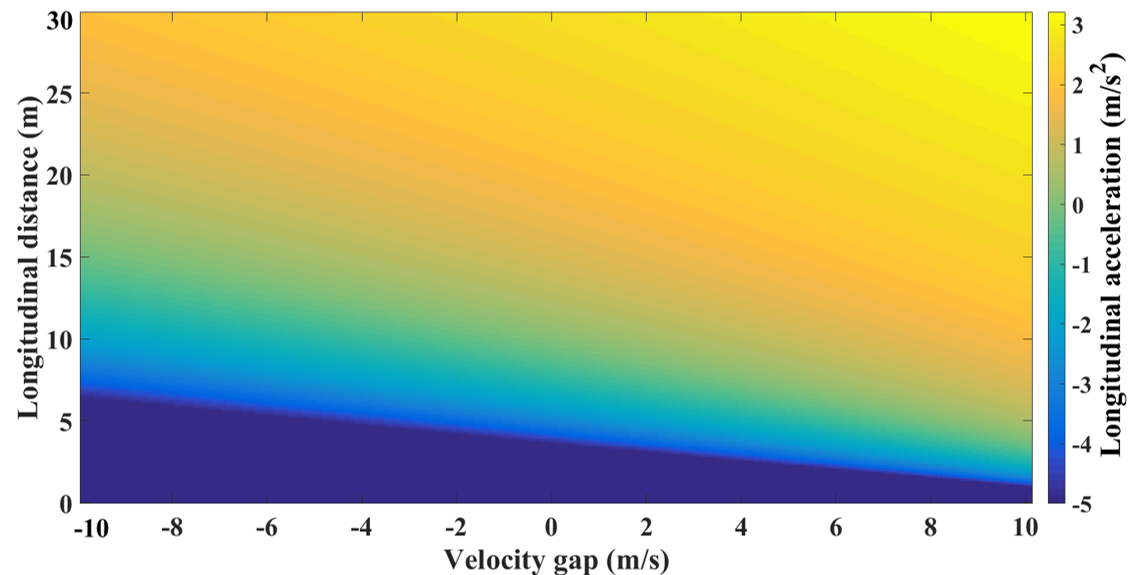}}
  \caption{Longitudinal Inter-vehicle potential function force}\label{fig:Inter-vehicle-force}
\end{figure}

The longitudinal inter-vehicle potential function is shown in Figure~\ref{fig:Inter-vehicle-function}, where $t_h$, $c_{ij}$, $x_{e}$ are set to $0.6 s$, 1, and $3 m$ respectively. Figure~\ref{fig:Inter-vehicle-force} shows  how $p_{x,ij}^b$ changes with velocity gap and the relative longitudinal distance, where $t_h$, $c_{ij}$, $x_{e}$ are set to $0.6 s$, 1, and $10 m$ respectively. $p_{x,ij}^b$ is lower than 0 (i.e. the repulsive force) if the distance between vehicles is smaller than $x_{e}-t_{h}\Delta v_{x,ij}$, while the magnitude of $p_{x,ij}^b$ increases sharply with the decreases in distance between vehicles. This specification meets the \textit{collision avoidance} principle from the safety perspective. When the velocity of the front vehicle is smaller than the velocity of the following vehicle, i.e. $\Delta v_{x,ij}<0$ and $x_{e}-t_{h}\Delta v_{ij}>x_e$, the following vehicle is supposed to maintain a larger safety distance under this situation. In contrast, if the distance is larger than $x_{e}-t_{h}\Delta v_{x,ij}$,  $p_{x,ij}^b$ ($>0$) increases slowly with the increases in distance and finally keeps the same velocity as the velocity of the front vehicle. As a result, vehicles tend to accelerate to reach the velocity of the front vehicle and maintain the gap of $x_{e}$, which satisfies the \textit{velocity matching} and \textit{flock centering} principles as well as the platooning requirements. Considering the domain of the potential function. $x_{e}-t_h\Delta v_{x,j}(t)$ should be larger than $0$. If it is not met, that means the velocity of the front vehicle is much larger than the following vehicle. The acceleration will be set as the largest allowed longitudinal acceleration. 

As to estimating the lateral inter-vehicle force $p_{x,ij}^b(y_i,y_j)$, $x_{e}-t_{h}\Delta v_{x,ij}$ in \eqref{eq:inter-vehicle-long} is replaced by $(y_e)_{ij}$, which is the constant lateral equilibrium distance. $(y_e)_{ij}$ is set as the lane width if vehicle $i$ and vehicle $j$ belong to different platoons or the longitudinal distance between them is smaller than 5 $m$. Otherwise, $(y_e)_{ij}=0$. The lateral inter-vehicle potential function is specified as follows:

  \begin{equation}
  p_{y,ij}^b(y_i,y_j)=c_{ij}(\ln(y_{ij})-\frac{(y_e)_{ij}\ln(y_e)_{ij}}{y_{ij}})
  \label{eq:inter-vehicle-late}
  \end{equation}

To be noted, if the distances between vehicles equal to $x_{e}$ and $(y_e)_{ij}$ respectively, and the vehicle pair has the same velocities $v_{x,i} = v_{x,j}$ (i.e. at equilibrium states), the inter-vehicle potential forces $p_{x,ij}^b$ and $p_{y,ij}^b$ equal to 0.

\subsubsection{Cross-section potential function} \label{subsection:cross-section}
The cross-section potential function is designed for the lane keeping and the lane changing behaviors. 
As for the lane keeping behavior, vehicles should stay within the lane  and be motivated to track the lane center of its target lane. As a result, the cross-section potential function is designed as a valley. The minimum and maximum values are set at the lane center of the desired lane and at the boundaries of the road segment respectively. The function values increase gradually when moving away from the lane center. The purpose of the maximum value $H$ is to keep vehicles within the drivable range of the road section.

As for the lane changing behavior, vehicles should be incentivized  to change lane by the inter-vehicle potential force $p_{y,ij}^b$. Vehicles are required to change lanes only when there is a vehicle in the same group in another lane, i.e. exerting a large force to change lane. As a result, a small positive value $h$ is set as a local maximum value at the lane markings between lanes, instead of the global maximum value. This can also prevent unnecessary lane changing behaviours.

\begin{figure}
  \centering
 \resizebox*{13cm}{!}{\includegraphics{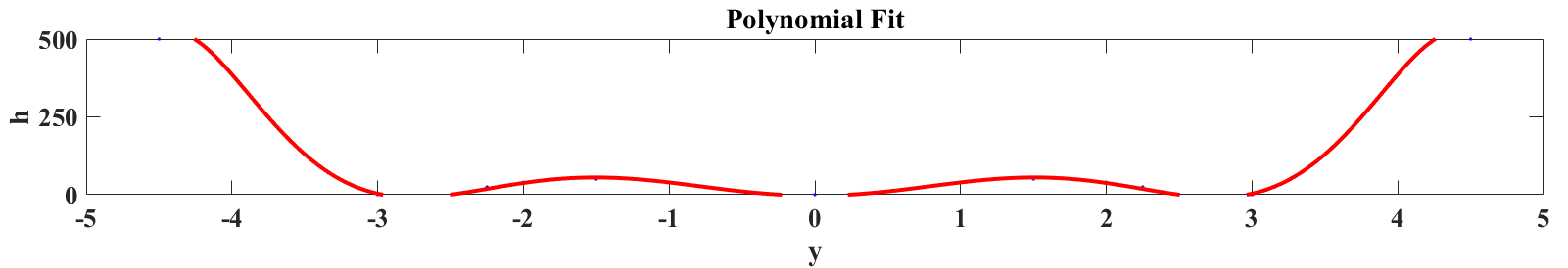}}
  \caption{Cross-section potential function with different fit methods}\label{fig:cross-section}
\end{figure}

From the aforementioned design, some feature points can be extracted. The heights of the feature points at central lines of all desired lanes equal to 0, while the heights of the feature points at the boundary between lanes are set to $h$. And the heights of the feature points at the road segment boundaries are set to $H$. $h$ and $H$ are 50 and 500 respectively. Interpolant fit, Polynomial fit and Fourier fit approaches are used to fit the cross-section potential function. Then, a simple scenario of three vehicles driving on a section with different initial states is designed to select the most suitable fitting method. The result shows that the Polynomial fit approach outperforms the others. The equation of the Polynomial fit and the cross-section potential function are shown in Eq. \eqref{eq:n2} and \eqref{eq:n4}. The figure of Polynomial fit is shown in Figure~\ref{fig:cross-section}.
  \begin{equation}
  f(y) = -0.0448y^8+1.738y^6-18.53y^4+59.36y^2\label{eq:n2}
  \end{equation}
  with the cross-section force: 
  \begin{equation}
  p_j^c=\frac{\partial f(y_j)}{\partial y_j}\label{eq:n4}
  \end{equation}

The design of the cross-section potential function also leads to a problem. If there is no vehicle in the perception zone, the lateral behavior of the vehicle will be controlled only by $p_j^c$. Under this situation, if the initial lateral position of the vehicle is not equal to the central line of the lane, this vehicle will exhibit oscillatory behaviour  in the “valley”. Therefore, a friction $f$ is introduced to avoid this problem:
  \begin{equation}
  p_j^f= f
  \end{equation}

\begin{figure}
  \centering
  \resizebox*{8cm}{!}{\includegraphics{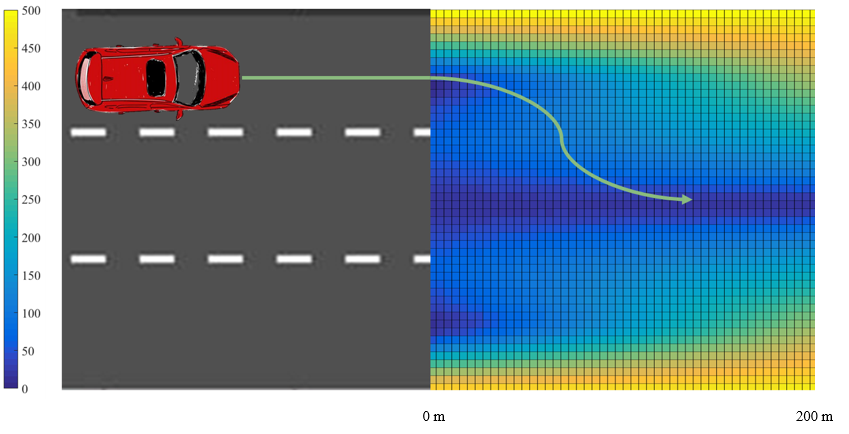}}
  \caption{Lane allocation by potential function}\label{fig:Lane-allocation}
\end{figure}

There are lane allocations in the approaching lanes, and CAVs need to pass the intersection using the correct lane. It is achieved by setting values of feature points. For instance, a vehicle on the left lane needs to use the middle lane to pass the intersection as shown in Figure~\ref{fig:Lane-allocation}. At the longitudinal point of 0 m, the feature point is the same as in Figure~\ref{fig:cross-section}, and all lanes can be used. In contrast, the heights of feature points are designed to trap the vehicle in the middle lane at the longitudinal point of 200 m. The heights of feature points change with different longitudinal positions, which is shown in Figure~\ref{fig:Lane-allocation}. This design can render the vehicle to move to the target lane gradually.

\subsubsection{Additional rules for formation dynamics}
Some additional rules are imposed to deal with the special situations. First, $p_{x,ij}^b$ and $p_{y,ij}^b$ between a CAV and other CAVs within the same platoon number or class should be different from $p_{x,ij}^b$ and $p_{y,ij}^b$ between a CAV and other vehicles. For instance, if two CAVs drive in parallel on different lanes of a road segment, they should change lane and form a platoon if they have the same platoon number or class, otherwise they should keep their current lanes. 
 Additionally, the sequence of each CAV may be set by the center controller. CAVs with the smaller sequence need to take over the CAVs with larger sequence. For a specific CAV, if another CAV is behind it and the lateral distance between them is smaller than the half width of the CAV exclusive lane, it is defined as a following CAV of the specific CAV. If a CAV detects that one of its following CAV has smaller sequence, a give-way behavior is activated. The give-way behavior is realized by two forces, i.e. a lateral force to push the ego vehicle to the right adjacent lane, and a longitudinal force to pull the ego vehicle to slow down. Finally, if a CAV is selected as a CAV platoon member and it follows another CAV with the same platoon or the same class in the desired sequence on the target lane, it will degrade to the car-following mode by implementing only $p_{x,ij}^b$ with the front vehicle.

When the front vehicle drives with the maximum velocity and its gap with the following vehicle is larger than the equilibrium distance,  the maximum velocity of the following vehicle is set to $v_x^{max}+v_x^{catch}$ until the gap between them meets the equilibrium distance, where $v_x^{catch}$ is a small positive value. This captures the approaching behavior in the free traffic flow. 

\section{Mathematical Model Properties} \label{section:calibration}
This section provides a mathematical analysis of the proposed flock-like model. The rationality of the car-following behavior is discussed in Subsection \ref{subsection:car-following-analysis}. And then subsection \ref{subsection:safety-analysis} proves the safety of the proposed flock-like model.

\subsection{Car-following model analysis} \label{subsection:car-following-analysis}
This section aims to discuss the rationality of the car-following behavior of the proposed model.  In \citet{wilson2011car}, the plausible behavior for general car-following models in the form of $a_{x,j} = f(\Delta x_j, \Delta v_{x,j}, v_{x,j})$ should have the properties as follows:
  \begin{equation}
  \frac{\partial a_{x,j}}{\partial \Delta x_j}>0
  \end{equation}
  \begin{equation}
  \frac{\partial a_{x,j}}{\partial \Delta v_{x,j}}>0
  \end{equation}
  \begin{equation}
  \frac{\partial a_{x,j}}{\partial v_{x,j}}<0
  \end{equation}
where $\Delta x_j$ and $\Delta v_{x,j}$ are calculated as in Eq. \eqref{eq:deltax} and \eqref{eq:deltav}.
  \begin{equation}
  \Delta x_j=x_{j-1}-x_j\label{eq:deltax}
  \end{equation}
  \begin{equation}
  \Delta v_{x,j}=v_{x,j-1}-v_{x,j}\label{eq:deltav}
  \end{equation}
  
The longitudinal acceleration of vehicle $j$ during platooning is determined by $p_{x,j(j-1)}^b$ and $p_j^d$, as shown in Eq.~\eqref{eq:10}.  The rational properties are proved from Eq.~\eqref{eq:11} to Eq.~\eqref{eq:13}. 
  \begin{equation}
  a_{x,j}(t)=c_{(j-1)j}(ln\text{\ensuremath{(\Delta x_{j}(t))-\frac{(x_{e}-t_h\Delta v_{x,j}(t))ln(x_{e}-t_h\Delta v_{x,j}(t))}{ \Delta x_{j}(t)}}})+p_j^d(v_{x,j})\label{eq:10}
  \end{equation}
  \begin{equation}
  \frac{\partial a_{x,j}(t)}{\partial v_{x,j}(t)}=-\frac{c_{(j-1)j}t_h}{\Delta x_{j}(t)}(ln(x_{e}-t_h\Delta v_{x,j}(t))+1)-\frac{F_{max}}{v_x^{max}}<0\label{eq:11}
  \end{equation}
  \begin{equation}
  \frac{\partial a_{x,j}(t)}{\partial \Delta v_{x,j}(t)}=\frac{c_{(j-1)j}t_h}{\Delta x_{j}(t)}(ln(x_{e}-t_h\Delta v_{x,j}(t))+1)>0\label{eq:12}
  \end{equation}
  \begin{equation}
  \frac{\partial a_{x,j}(t)}{\partial \Delta x_{j}(t)}=\frac{c_{(j-1)j}}{\Delta x_{j}(t)}(1+\frac{(x_{e}-t_h\Delta v_{x,j}(t))ln(x_{e}-t_h\Delta v_{x,j}(t))}{\Delta x_{j}(t)})>0
  \label{eq:13}
  \end{equation}

\subsection{Emergency stopping scenario analysis}\label{subsection:safety-analysis}
This subsection investigates the safety behavior of the model when platoon leader suddenly decelerates from the desired velocity to a full stop, which is one of the most common test scenarios for longitudinal safety. The emergency stropping process is analyzed using discretization. The longitudinal distance can be calculated as follows. 

$x_{j}(t)$ is a basic parameter in safety analysis, obtained by Eq. \eqref{eq:16}. Eq. \eqref{eq:17} evloves from Eq.~\eqref{eq:16}. The spacial gap is a critical parameter for safety analysis, calculated using $x_{j-1}(t)-x_{j}(t)$, as shown in Eq.~\eqref{eq:18}.
  \begin{equation}
  x_{j}(t)=x_{j}(0)+\Delta t\sum_{m=1}^{t-1}v_{x,j}(m)+\frac{\Delta tv_{x,j}(0)}{2}+\frac{\Delta tv_{x,j}(t)}{2}\label{eq:16}
  \end{equation}
  \begin{equation}
  x_{j}(t)=x_{j}(0)+\Delta t\left(tv_{x,j}(0)+\sum_{m=0}^{t-1}\left(t-\frac{1}{2}-m\right)a_{x,j}(m)\right)\label{eq:17}
  \end{equation}
  \begin{equation}
  \frac{\Delta x_{j}(t)}{\Delta t}=\frac{x_{e}}{\Delta t}+\sum_{m=0}^{t-1}\left(t-\frac{1}{2}-m\right)\left(a_{x,j-1}(m)-a_{x,j}(m)\right)\label{eq:18}
  \end{equation}
Introducing $\Delta x_{j}(t-1)$ to Eq.~\eqref{eq:18}, the relationship between $\Delta x_{j}(t)$ and $\Delta x_{j}^(t-1)$ is shown as in Eq.~\eqref{eq:19}.
  \begin{equation}
  \frac{\Delta x_{j}(t+1)-\Delta x_{j}(t)}{\Delta t}=\Delta t\sum_{m=0}^{t-1}\left(a_{x,j-1}(m)-a_{x,j}(m)\right)+\frac{\Delta t}{2}\left(a_{x,j-1}(t)-a_{x,j}(t)\right)\label{eq:19}
  \end{equation}
Since $v_{x,j-1}(0)=v_{x,j}(0)$, the first part of the left side, $\Delta t\sum_{m=0}^{t-1}\left(a_{x,j-1}(m)-a_{x,j}(m)\right)$, is the velocity gap at time $t-1$. Eq.~\eqref{eq:19} can be transformed into Eq.~\eqref{eq:20}.
  \begin{equation}
  \Delta x_{j}(t+1)-\Delta x_{j}(t)=\text{\ensuremath{\Delta v_{x,j}(t-1)}}\Delta t+\frac{{\Delta t}^2}{2}\left(a_{x,j-1}(t)-a_{x,j}(t)\right)\label{eq:20}
  \end{equation}
The relationship between the space headway when the vehicle has a full stop and the space headway at the initial state is shown in Eq.~\eqref{eq:21}, which is the summation of Eq.~\eqref{eq:20} from $t=0$ to $t=t_j^{end}$. $t_j^{end}$ indicates the time when vehicle $j$ has a full stop.
  \begin{equation}
  x_{e}-\Delta x_{j}(t_j^{end})=\sum_{m=1}^{t_j^{end}}\Delta v_{x,j}(m)\Delta t\label{eq:21}
  \end{equation}
With respect to the car-following behavior, it is dangerous for a vehicle if it has to slow down with the largest allowable deceleration to avoid collision. Under this situation, the space headway when the vehicle has to fully stop can be indicated as:
  \begin{equation}
  x_{e}-\Delta x_{j}(t_j^{end})<\Delta v_x^{max}\frac{v_x^{max}}{a_x^{min}}\label{eq:22}
  \end{equation}
The space headway of the final state, $\Delta x_{i}^{t_j^{end}}$, needs to be larger than 0 to guarantee safety, therefore:
  \begin{equation}
  x_{e}>\Delta v_x^{max}\frac{v_x^{max}}{a_x^{min}}\label{eq:23}
  \end{equation}
In the initial state, all vehicles are operated at the desired velocity with the space headway of $x_{e}$. Then, the front vehicles start to slow down with the largest deceleration, while the following vehicles still run with the desired velocity, until the space headway reaches the threshold $(p_{x,ij}^b)^{-1}(a_x^{min})$. $(p_{x,ij}^b)^{-1}(a)$ is the reverse function of $p_{x,ij}^b$ with $\Delta v_{x,ij}=0$. 
 $(p_{x,ij}^b)^{-1}(a_{x}^{min})$ is the space headway that leads to the largest deceleration. And the absolute value of the velocity gap $\Delta v_{x,ij}$ meets the maximum value when the space headway meet  $(p_{x,ij}^b)^{-1}(a_{x}^{min})$.
 The behaviors are illustrated in Eq.~\eqref{eq:24} and Eq.~\eqref{eq:25}.
  \begin{equation}
  \Delta v_x^{max}=a_x^{min}t_{con}\label{eq:24}
  \end{equation}
where $t_{con}$ indicates the time period of vehicle $j$ driving at the constant largest velocity $v_x^{max}$ after the deceleration of the front vehicle $i$.
  \begin{equation}
  \frac{a_x^{min}{t_{con}}^{2}}{2}=(p_{x,ij}^b)^{-1}(a_x^{min})-x_{e}\label{eq:25}
  \end{equation}
According to Eq.~\eqref{eq:24} and Eq.~\eqref{eq:25}, $\Delta v_x^{max}$ can be indicated as:
  \begin{equation}
  \Delta v_x^{max}=\sqrt{2a_x^{min}((p_{x,ij}^b)^{-1}(a_x^{min})-x_{e})}\label{eq:26}
  \end{equation}
By introducing Eq.~\eqref{eq:26} to Eq.~\eqref{eq:23}, the safety condition can be expressed as: 
  \begin{equation}
  x_{e}>v_x^{max}\sqrt{\frac{2((p_{x,ij}^b)^{-1}(a_x^{min})-x_{e})}{a_x^{min}}}\label{eq:27}
  \end{equation}

If this condition of Eq. \eqref{eq:27} is respected, the platoon safety  can be guaranteed under the scenario that the leader has a sharp deceleration from the desired velocity to a full stop. The space headway is always larger than a positive number, which is further proved in the following numerical experiments, as shown in Figure~\ref{fig:result-xe=3} and Figure~\ref{fig:result-xe=30}.

\section{Numerical experiments} \label{section:evolution}
In this section, numerical experiments are conducted to demonstrate the workings of the proposed flock-like model. The experiments are designed in \textbf{Subsection Experiments Design}. Then, the experiment results are presented in \textbf{Subsection Numerical Experiments Result}, followed by the discussions and conclusions on experiment results in \textbf{Subsection Discussion}.

\subsection{Experiments design}\label{subsection:Experiments-design}
The numerical experiments aim to verify the behaviors of the proposed flock-like model. The scenarios are designed to demonstrate the typical behaviors of car following, lane changing, platooning and splitting, which can be accordingly  categorized into the longitudinal scenarios, two-dimensional scenarios and the comprehensive scenarios. The longitudinal scenario aims to evaluate the car-following behavior within a platoon (see Platoon emergency stopping test in table \ref{tab:scenario-design}). The two-dimensional scenarios are proposed to verify the lane changing, splitting and platoon forming behaviors for a individual vehicle and further for multiple vehicles (see Single platoon forming test, Multi-platoons forming test, Single vehicle lane changing test, and Multi-vehicles separation test in Table \ref{tab:scenario-design}). The comprehensive scenarios test the performance of forming a single platoon and multiple platoons under the mixed traffic environment (see Vehicles not in a group
form in a platoon test and Vehicles in a group form in a platoon test in Table \ref{tab:scenario-design}).  

\begin{figure*}
  \centering
  \subfigure[Illustration of test 1]{
  \resizebox*{6cm}{!}{\includegraphics{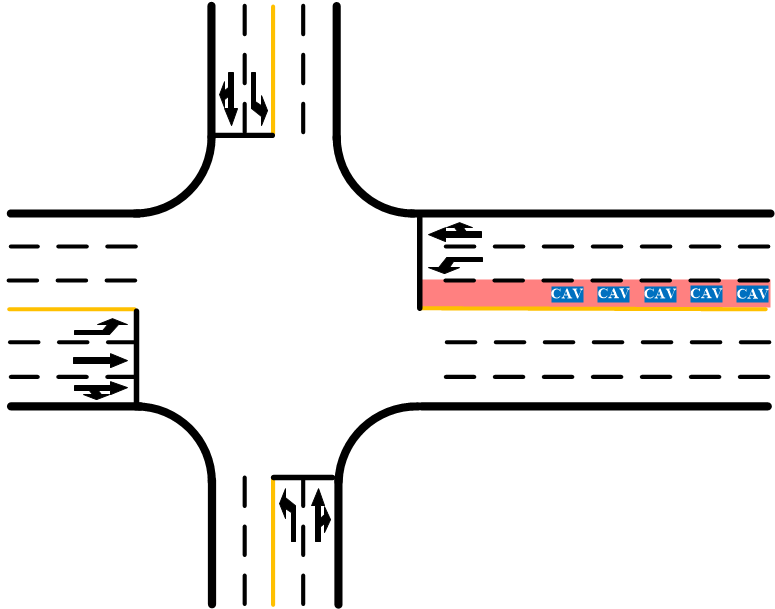}}}
  \subfigure[Illustration of test 2] { 
\resizebox*{6cm}{!}{\includegraphics{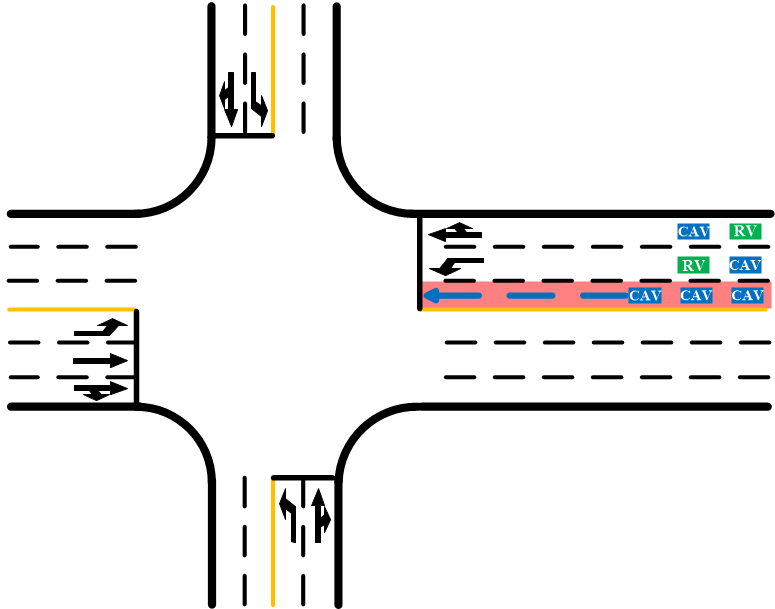}} 
} 
\subfigure[Illustration of test 3]{
  \resizebox*{6cm}{!}{\includegraphics{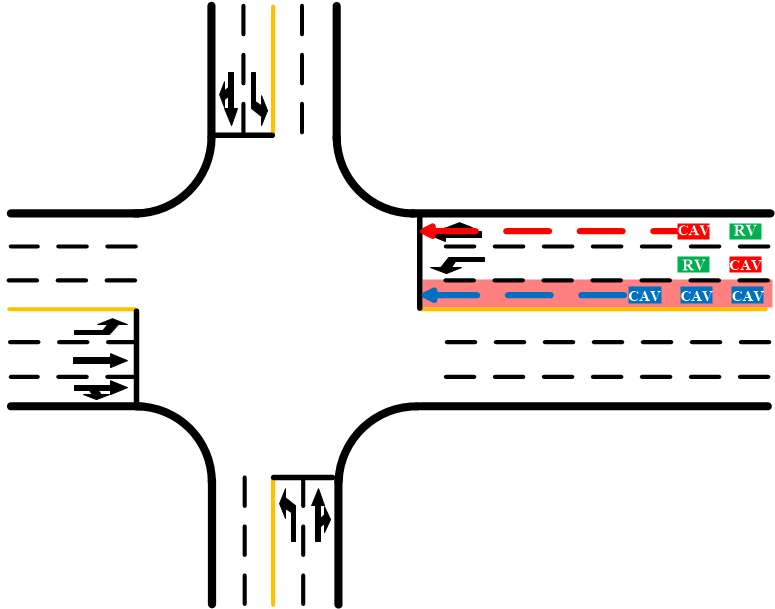}}}
  \subfigure[Illustration of test 4] { 
\resizebox*{6cm}{!}{\includegraphics{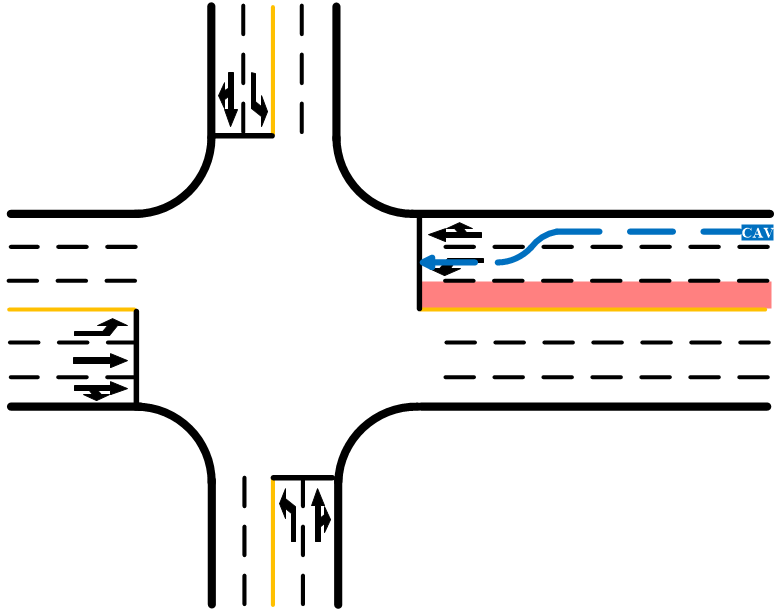}}     
} 
\subfigure[Illustration of test 5] { 
\resizebox*{6cm}{!}{\includegraphics{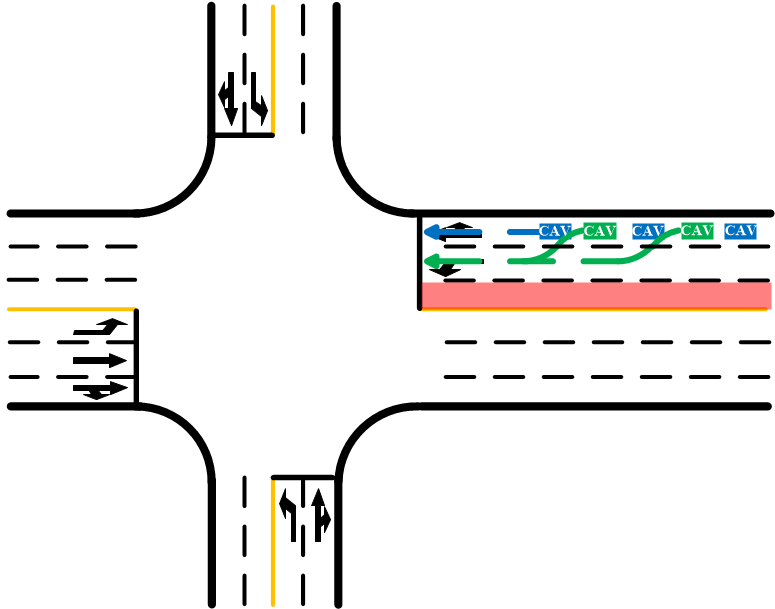}}     
}
\quad
\subfigure[Illustration of test 6] { 
\resizebox*{6cm}{!}{\includegraphics{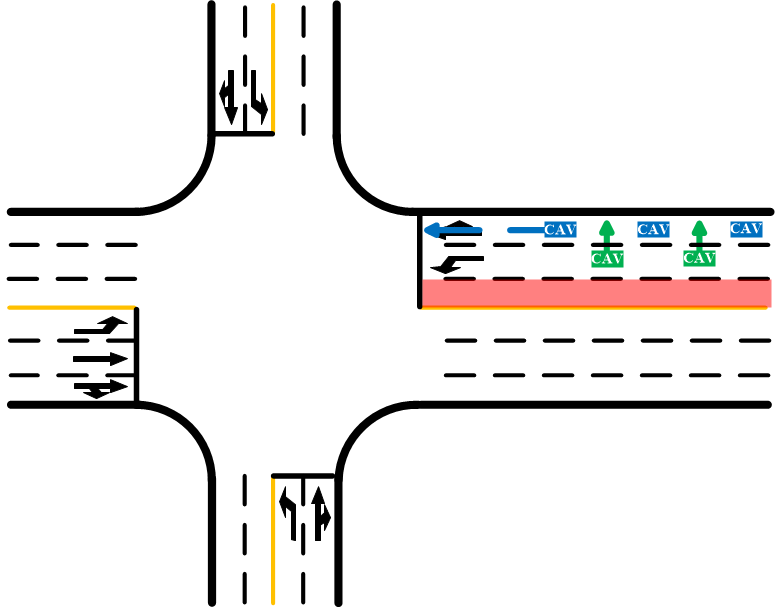}}     
}
\subfigure[Illustration of test 7]{
  \resizebox*{6cm}{!}{\includegraphics{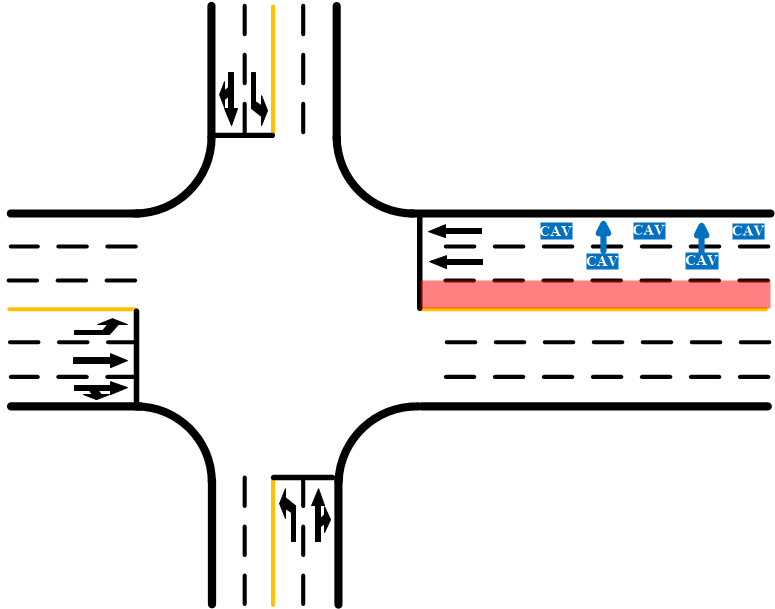}}}
\quad
\caption{Illustration of test scenarios}\label{fig:scenarios}  
\end{figure*}

All test scenarios are illustrated in Figure~\ref{fig:scenarios}. The pink lane in this figure is the CAV dedicated lane, and vehicles in a same platoon have the same color. 
The length of the test road is set as 600 m with the link velocity limitation of 20 m/s. An intersection is located on the link at the longitudinal position of 400 m. The numerical study is simulated in MATLAB platform for a period of  30 s. 

In the numerical experiments, the maximum longitudinal acceleration, the maximum longitudinal deceleration, the maximum lateral acceleration, the maximum longitudinal velocity and the maximum lateral velocity are set to be 3 $m/s^2$, -5 $m/s^2$, 2 $m/s^2$, 20 $m/s$ and 1 $m/s$ respectively.

\begin{table}[!ht]
	\tbl{Numerical experiments design}{

	\centering
		\begin{tabular}{p{30pt}p{80pt}p{100pt}p{150pt}}\hline
			Test number & Class & Scenario & Purpose\\\hline
			1 & Longitudinal Scenario & Platoon emergent stopping & Car following in accelerating and decelerating \\
			2 & Comprehensive Scenario & Single platoon forming  & Cooperative merging \\
			3 & Comprehensive Scenario & Multi-platoons forming & Management and cooperative merging \\
			4 & Two-dimensional Scenario & Single vehicle lane changing & Lane changing with different values of parameters \\
			5 & Two-dimensional Scenario & Multi-vehicles separation & Lane changing and platoon separation with different values of parameters \\
			6 & Two-dimensional Scenario & Vehicles not in a group form in a platoon & Lane changing and platoon forming with different values of parameters \\
			7 & Two-dimensional Scenario & Vehicles in a group form in a platoon & Lane changing and platoon forming with different values of parameters \\\hline
		\end{tabular}}
	\label{tab:scenario-design}
\end{table}

\subsection{Numerical experiments result}\label{subsection:using-cases}
\begin{figure*}
  \centering
  \subfigure[Vehicle trajectories]{
  \resizebox*{6.5cm}{!}{\includegraphics{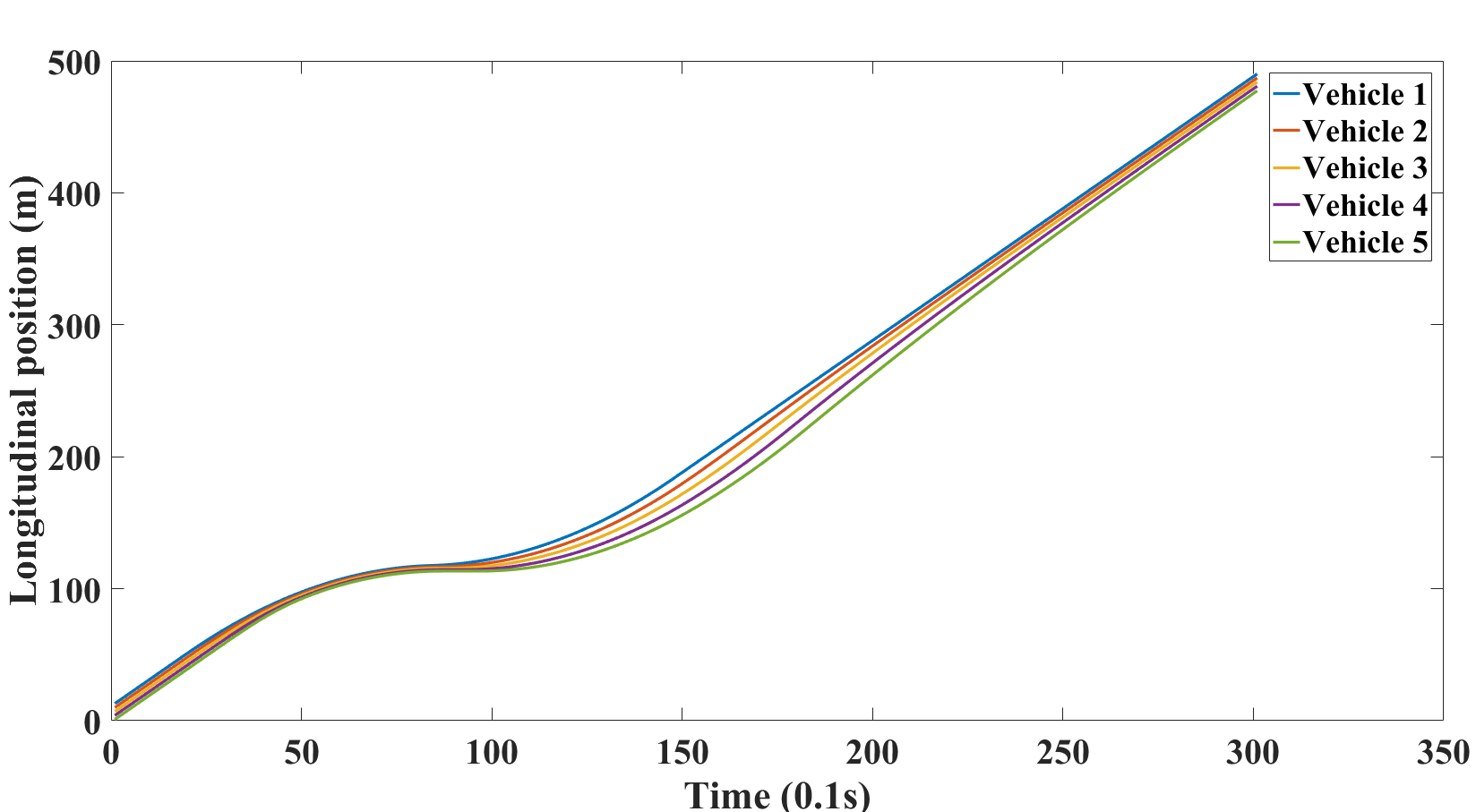}}}
  \subfigure[Space headway-t figure] { 
\resizebox*{6.5cm}{!}{\includegraphics{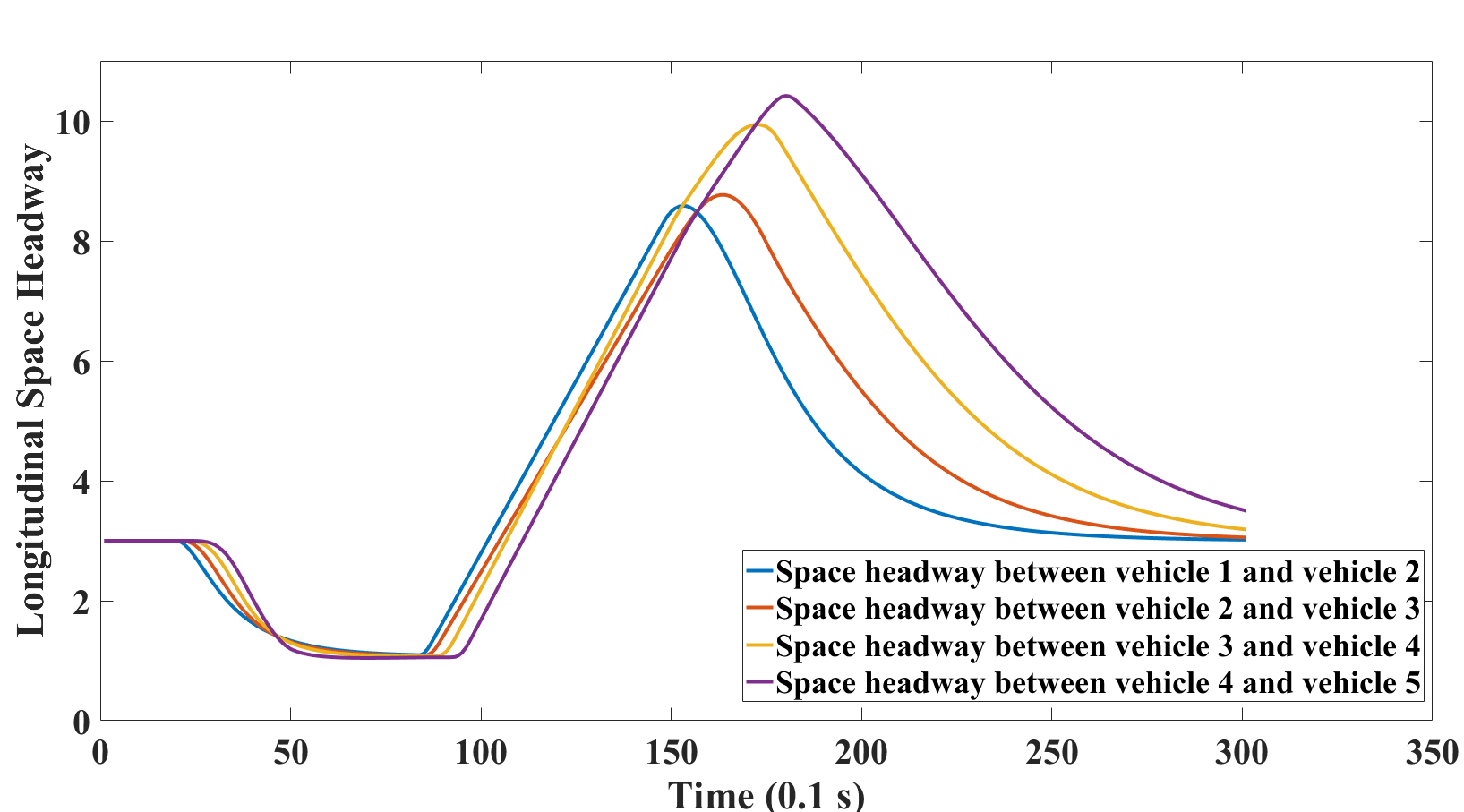}}     
} 
\quad
\subfigure[Longitudinal velocities figure] { 
\resizebox*{6.5cm}{!}{\includegraphics{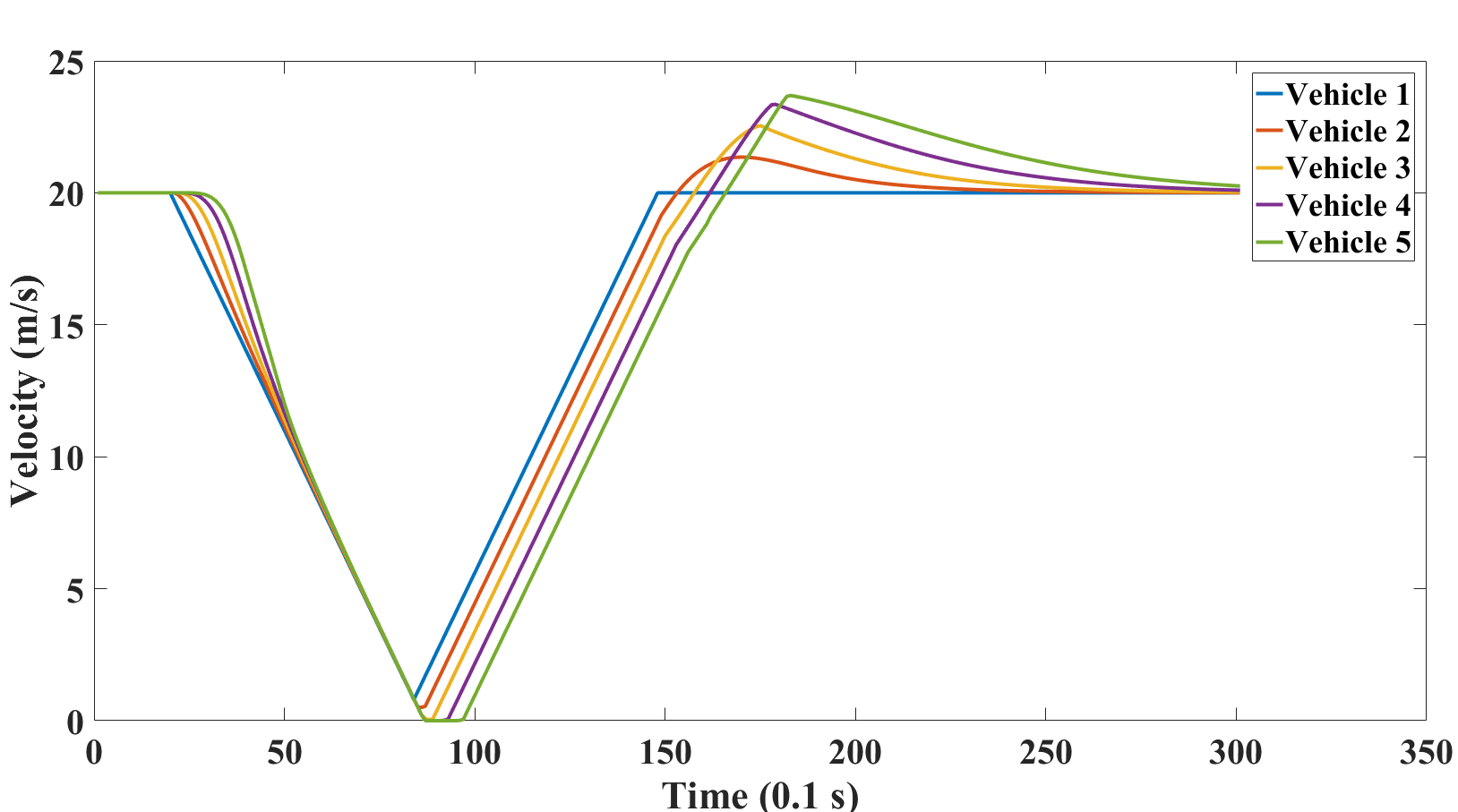}}    
} 
\subfigure[Longitudinal accelerations figure] { 
\resizebox*{6.5cm}{!}{\includegraphics{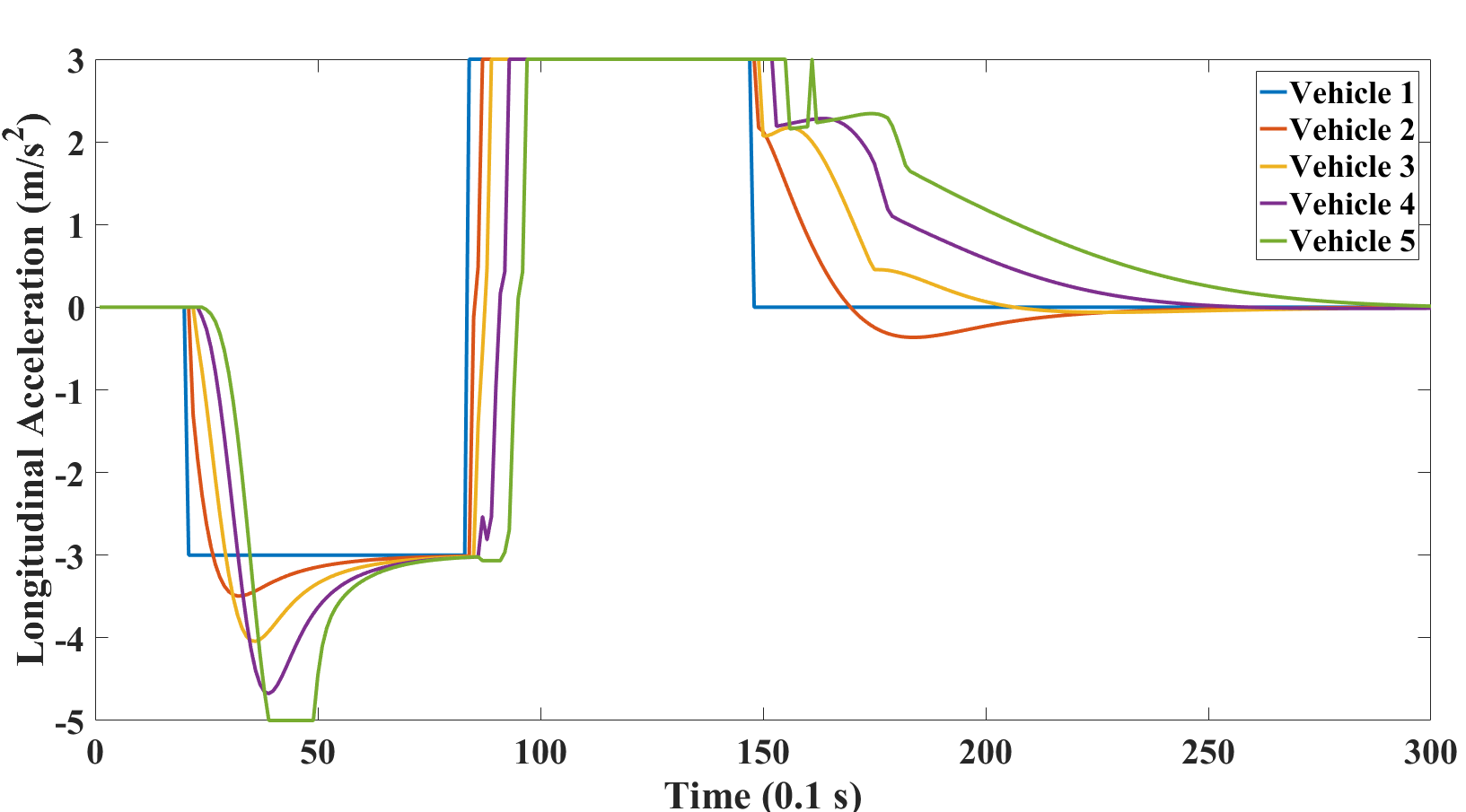}}  
}
  \caption{Result with $x_e$ equals to 3}\label{fig:result-xe=3}  
\end{figure*}

\begin{figure*}
  \centering
  \subfigure[Vehicle trajectories] {
  \resizebox*{6.5cm}{!}{\includegraphics{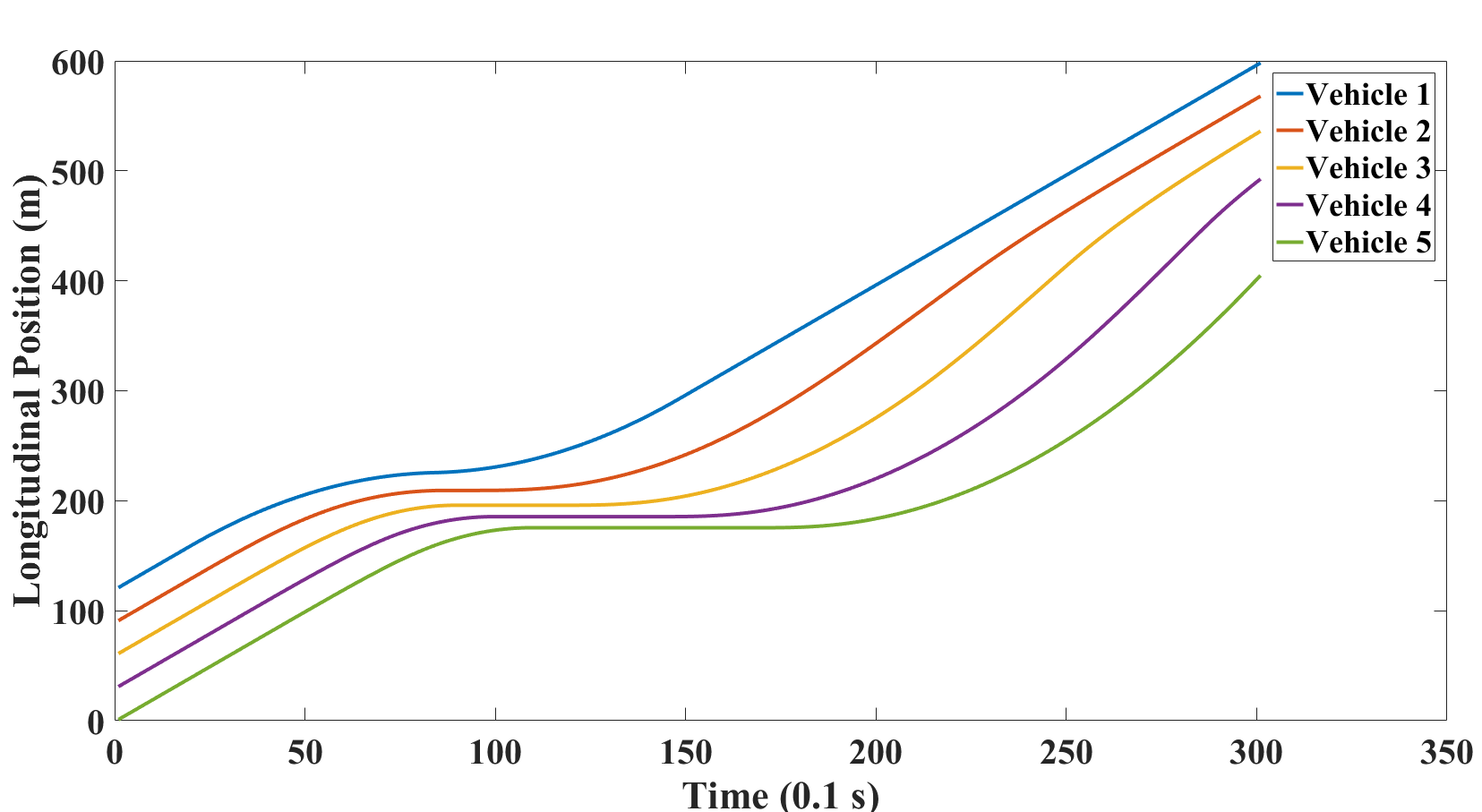}}}
  \subfigure[Space headway-t figure] { 
\resizebox*{6.5cm}{!}{\includegraphics{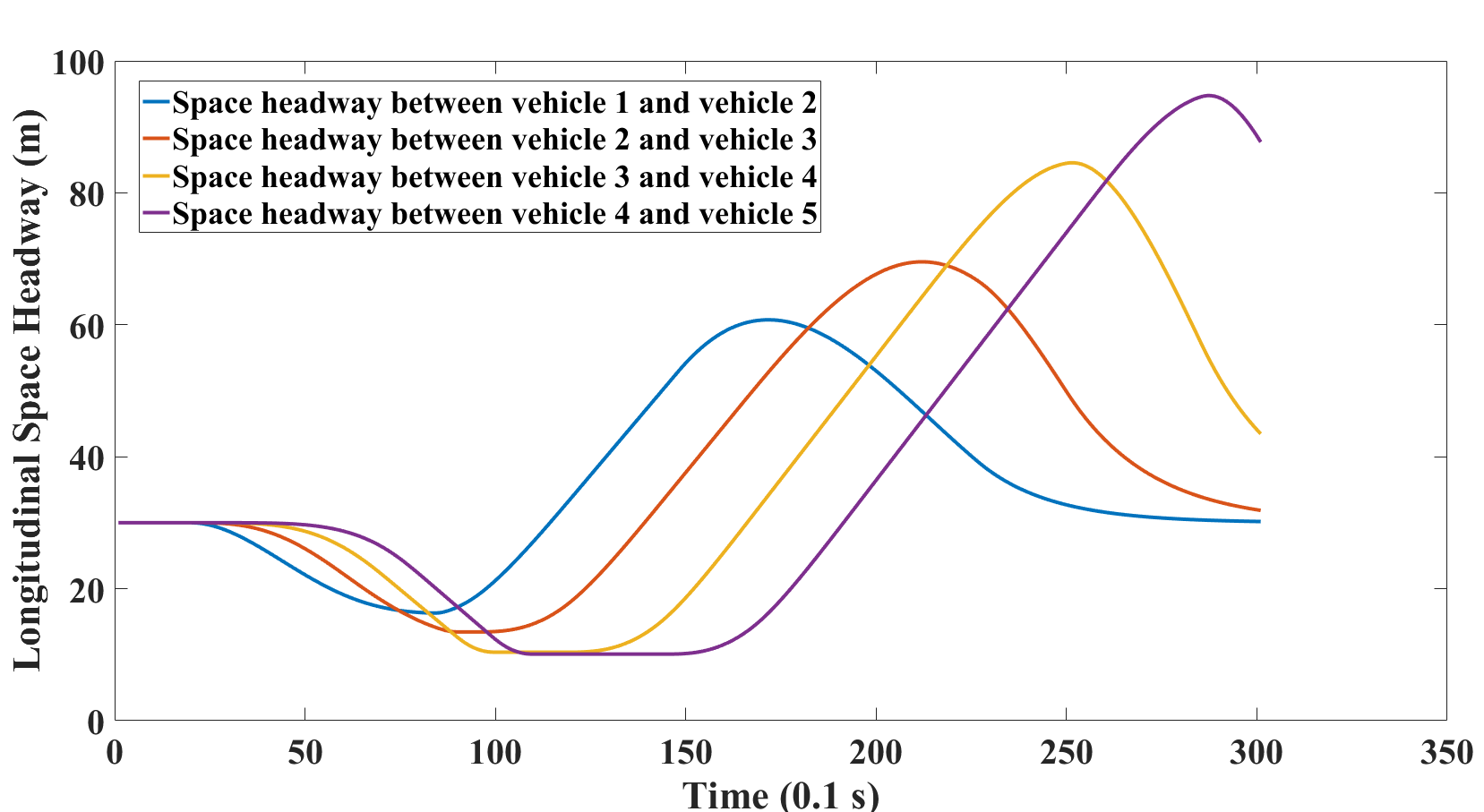}}     
} 
\quad
\subfigure[Longitudinal velocities figure] { 
\resizebox*{6.5cm}{!}{\includegraphics{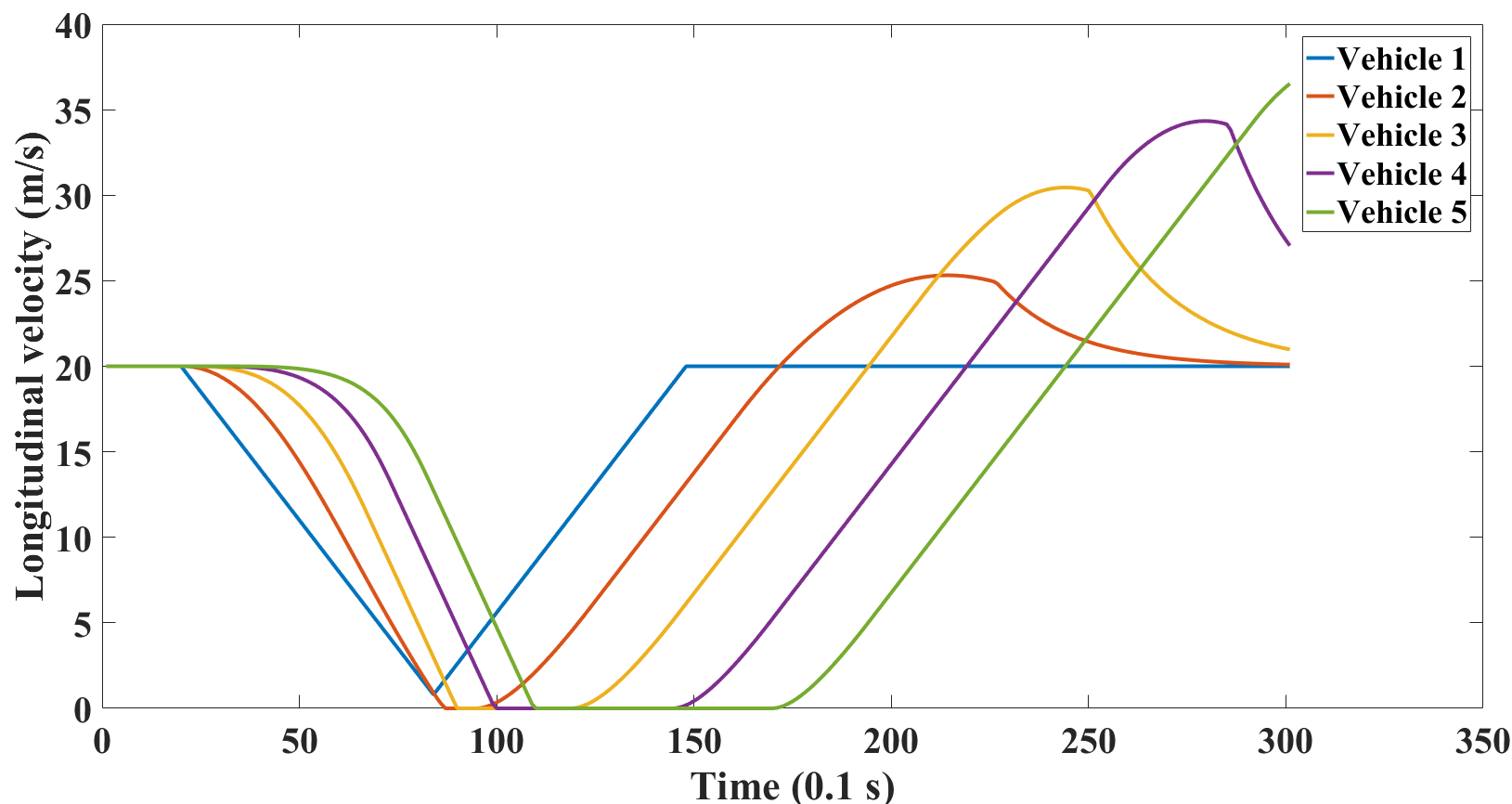}}     
}
\subfigure[Longitudinal accelerations figure] { 
\resizebox*{6.5cm}{!}{\includegraphics{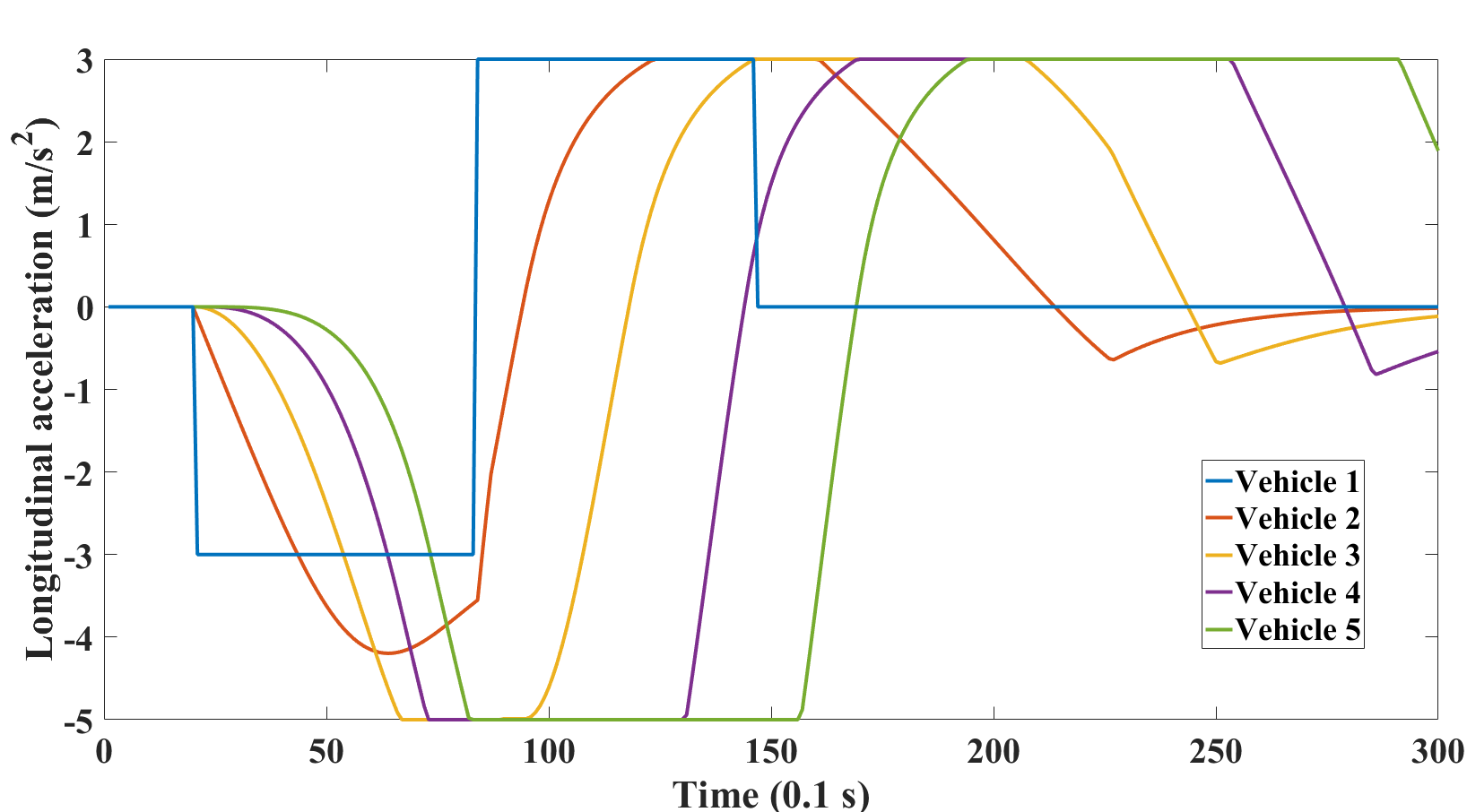}}   }
  \caption{Result with $x_e$ equals to 30}\label{fig:result-xe=30}
\end{figure*}

For all tests, the model can run in real-time. In Platoon emergency stopping test, the car-following behavior is shown to be safe by using the "emergent stopping" scenario, as shown in Figure \ref{fig:result-xe=3}. A platoon composed of five CAVs are proposed to test this scenario. At first, all vehicles travel with the desired velocity, and the distances between each pair of adjacent vehicles are equal to the equilibrium distance. The first vehicle has an emergency brake at around 20 s and decelerates until 0 m/s, and then it accelerates to the maximum velocity again, as shown in the V-t subfigure in Figure~\ref{fig:result-xe=3}. According to the safety analysis discussed in Eq.~\eqref{eq:27}, safety can be guaranteed if the equilibrium distance is larger than a threshold value. Different parameter values of $x_{e}=3 $ m and $x_{e}=30$ m are tested respectively, and the corresponding trajectories, space headway, velocity and acceleration of vehicles in the platoon under $x_{e}=3$ m and $x_{e}=30$ m are presented in Figure~\ref{fig:result-xe=3} and Figure~\ref{fig:result-xe=30}. All vehicles respect the safety driving requirements, although the minimum gap is smaller than the propagation of the platoon and the velocity fluctuations are amplified. However, in the space headway-t subfigures of Figure~\ref{fig:result-xe=3} and Figure~\ref{fig:result-xe=30}, the minimum gap is finally stable at around 1 m and 10 m respectively, in spite of the fluctuating platoon propagation, which proves the conclusion of Subsection \ref{subsection:safety-analysis} that the final longitudinal gap will always be larger than a value. And this value can be adjust by using suitable parameters such as $x_{e}$. Additionally, by comparing Figure~\ref{fig:result-xe=3} and Figure~\ref{fig:result-xe=30}, it can be found that larger values of $x_{e}$ also lead to larger minimum space headways, while lower values of $x_{e}$ result in faster convergence.

\begin{figure}
  \centering
  \resizebox*{10cm}{!}{\includegraphics{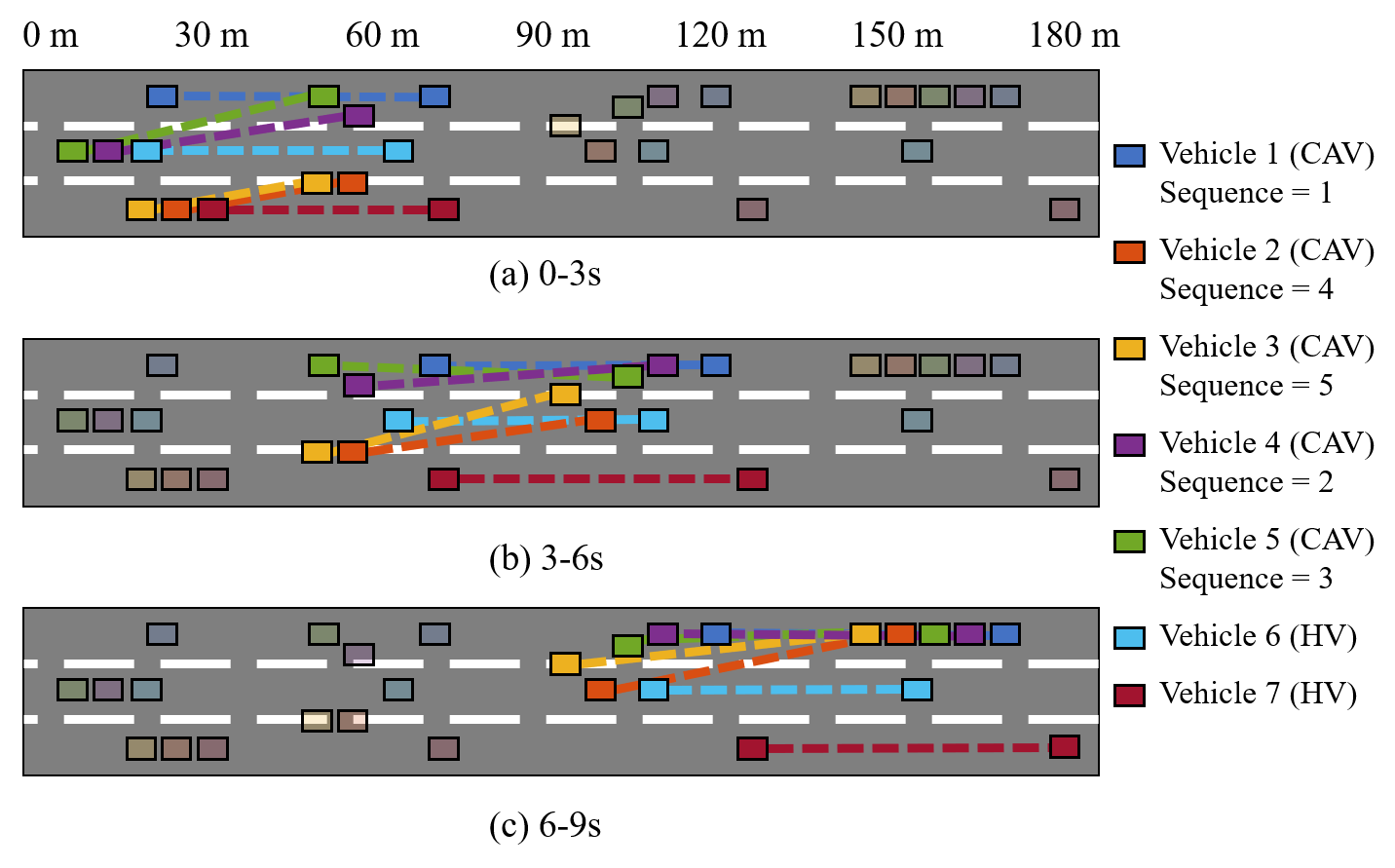}}
  \caption{Test result of Single platoon forming test}.\label{fig:multi-vehicles}
\end{figure}

Hereinafter, the comprehensive experiments are demonstrated in two scenarios under the mixed traffic environment, i.e. the single platoon (Single platoon forming test) and multi-platoon (Multi-platoons forming test). Seven vehicles composed of five CAVs and two HVs are initially set on different lanes of the test link. In the first experiment, the five CAVs form a platoon from different lanes. The desired velocity of all vehicles is 20 m/s. The longitudinal and lateral trajectories are shown in Figure~\ref{fig:multi-vehicles}. It is clear that the vehicle sequence is adjusted in this figure, e.g. vehicle 4 and 5 take over vehicle 2 and 3. Providing CAV exclusive lanes, this performance is beneficial to the improvement in lane utilization under the adaptive control at the urban network. The proposed model can complete the platoon sequence adjustment around 90 m and the platoon formation around 150 m to the adjusting-start-line.

\begin{figure}
  \centering
  \resizebox*{10cm}{!}{\includegraphics{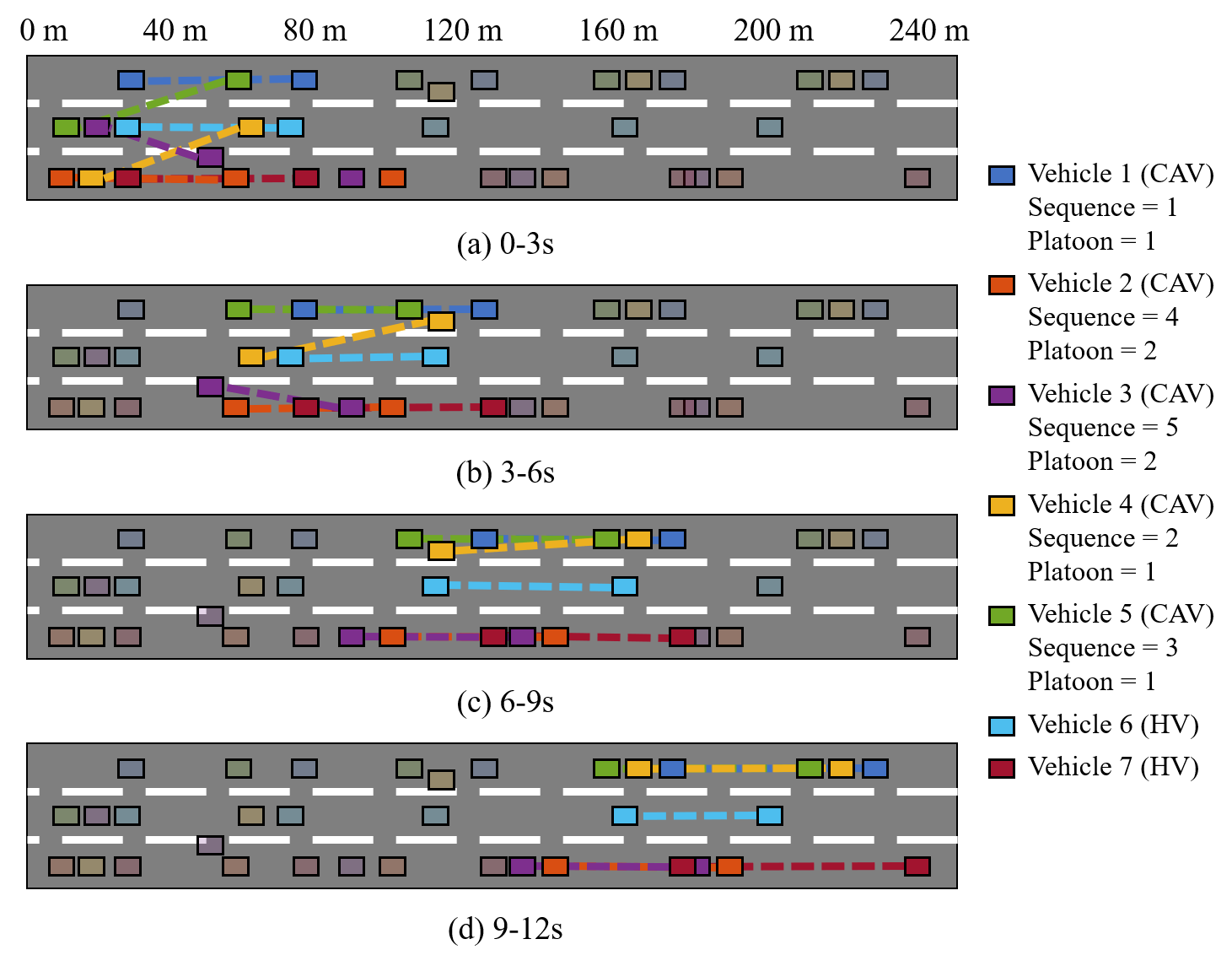}}
  \caption{Test result of Multi-platoons forming test.}\label{fig:multi-platoons}
\end{figure}

The multi-platoon scenario is tested considering two platoons. Platoon 1 is composed of vehicle 1, 4 and 5, and platoon 2 consists of vehicle 2 and 3. Vehicle 1 and 2 are the leader of platoon 1 and platoon 2 respectively. The desired velocity of platoon 1 is set to 20 m/s, while the desired velocity of platoon 2 is 18 m/s. More than four lane changing and overtaking behaviours are applied to form platoons. The results of the multi-platoon scenario are shown in Figure~\ref{fig:multi-platoons}. The platoon formation finishes within 150 m. HVs keep the initial driving strategy because they are not influenced by the following CAVs. If CAVs form platoons passively without lane changing (for instance, forming platoons according to the communication range, which is widely used in the existing platoon-based urban traffic control models \citep{Cooper2016,Jin2013}), three platoons are generated respectively on their initial lanes, i.e. vehicle 1, vehicle 3 and 5, and vehicle 2 and 4. In this way, vehicle 4 and 5 can not reach the desired velocity, and even no platoon is formed if they have different routes. On the other hand, if CAVs form platoons passively with considering lane changing, there are still three platoons, that is, vehicle 1, vehicle 2 and 3, and vehicle 4 and 5. Since vehicle 2 and vehicle 4 are in the same lane, the platoon composed of vehicle 4 and 5 has to follow platoon 2. Vehicle 4 and vehicle 5 can not reach the desired velocity. If they have different routes, vehicle 4 and vehicle 5 possibly have to wait for another signal cycle to pass the intersection, causing additional travel delay. Through Single platoon forming test and Multi-platoons forming test, the proposed flock-like model can incorporate the vehicle sequence adjustment and the reasonable two-dimensional motions, both of which are able to better utilize the urban temporal-spatial resource under the adaptive control at intersections. To conclude, the proposed flock-like model has a good performance in the numerical experiments.

\subsection{Numerical experiments discussion}\label{subsection:discussion}
\begin{figure*}
  \centering
  \subfigure[Vehicles in adjacent lane] {
  \resizebox*{6.5cm}{!}{\includegraphics{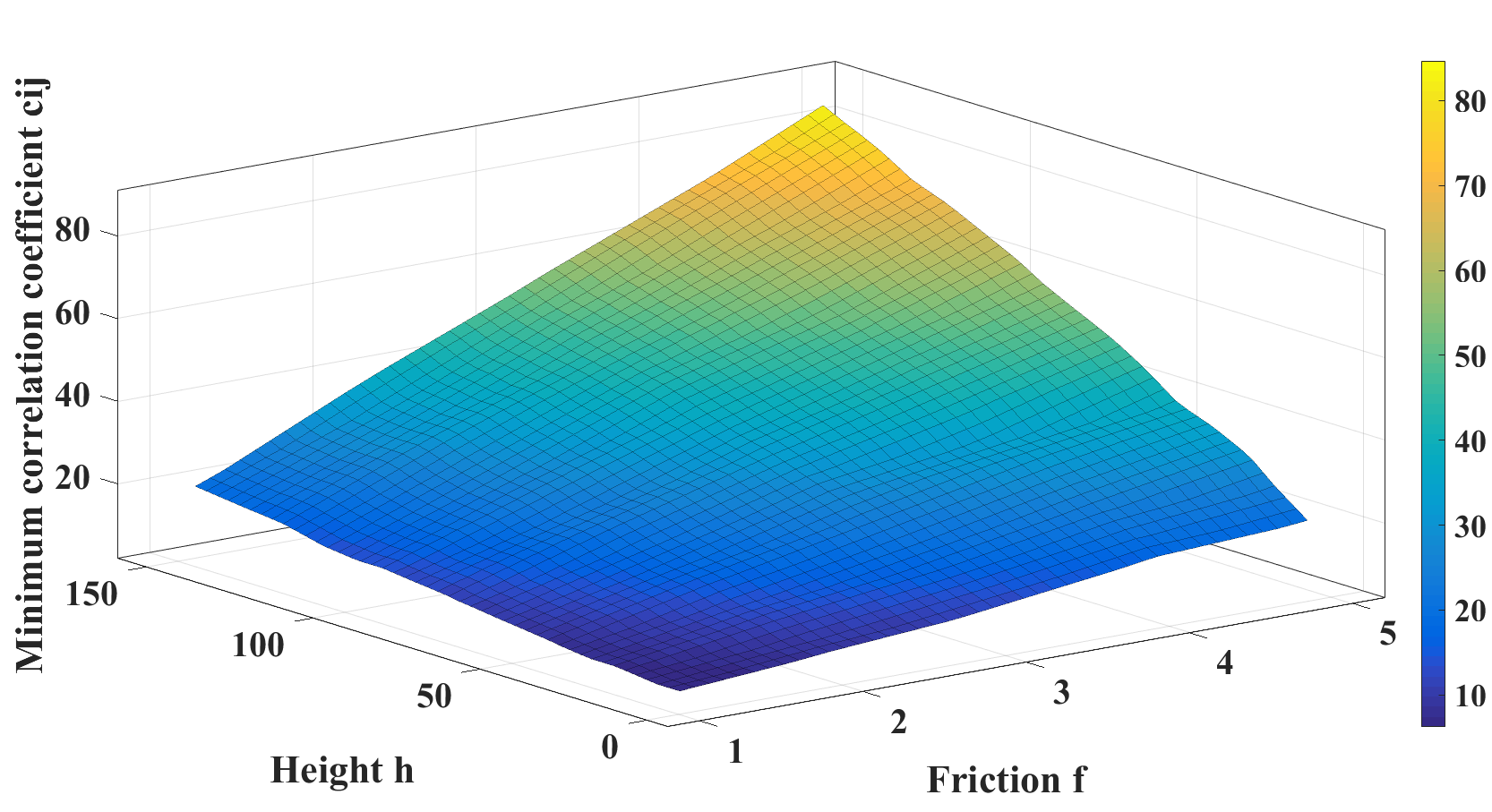}}}
  \subfigure[Vehicles separated by one lane] {
  \resizebox*{6.5cm}{!}{\includegraphics{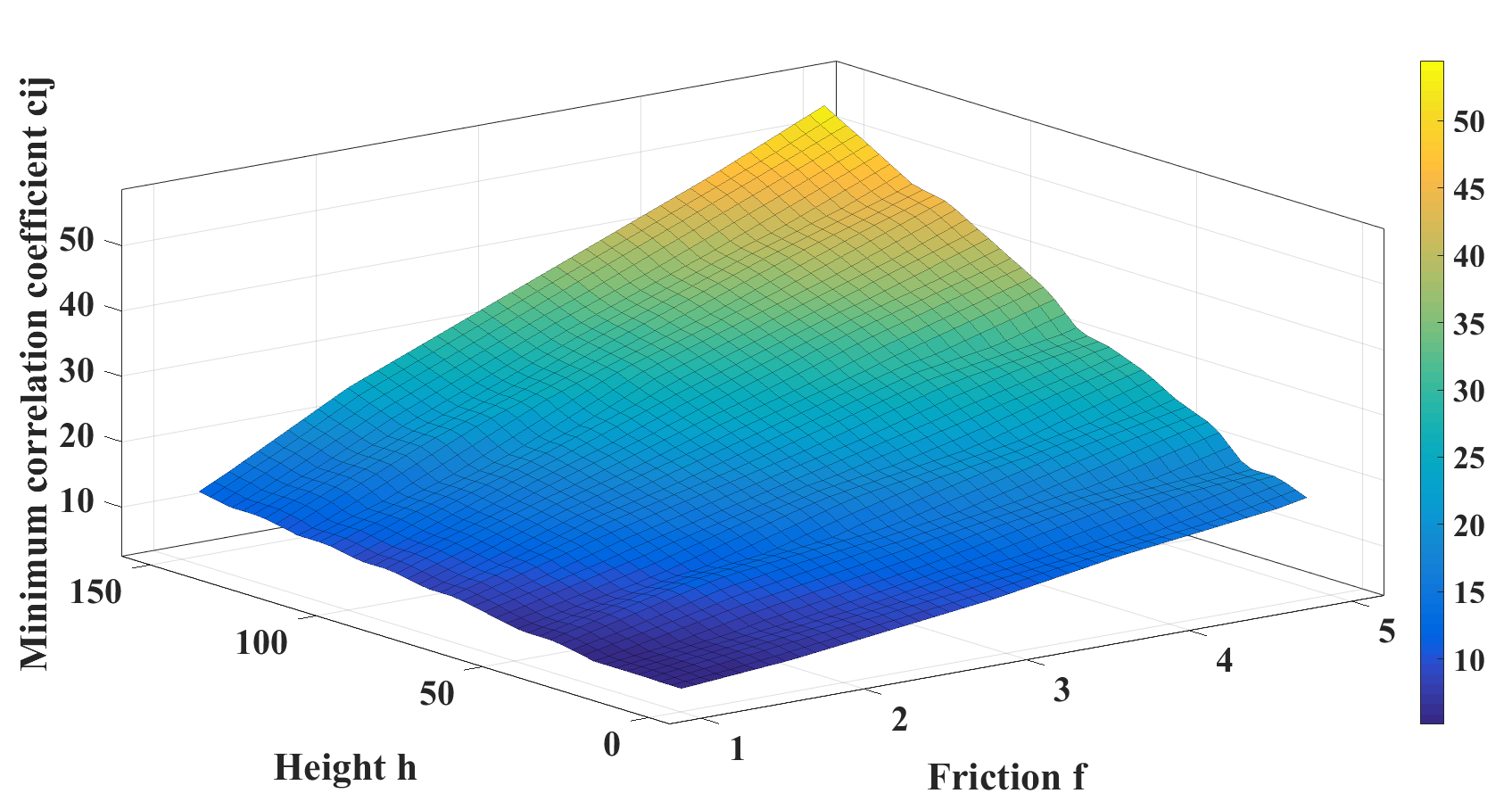}}}
  \caption{Reasonable range of minimum correlation coefficient}\label{fig:cali-result}
\end{figure*}

In this section, analysis on several numerical experiments are discussed. As mentioned above, Platoon emergency stopping test shows that the platoon safety can be guaranteed, and Single platoon forming test and Multi-platoons forming test show the value of the proposed algorithm. However, there are several parameters which may influence the performance of the proposed algorithm.
 The reasonable parameter ranges still need to be explored. Several two-dimensional scenarios including individual vehicle scenarios and multi-vehicle scenarios are proposed to explore it (Single vehicle lane changing test, Multi-vehicles separation test, Vehicles not in a group form in a platoon test, and Vehicles in a group form in a platoon test). 
 The objective of the single vehicle test (Single vehicle lane changing test) is to change to the target lane in the vicinity of intersection. The test vehicle is initially on the right lane, and intends to use the left lane to pass the intersection under appropriate parameter values. Two-dimensional scenarios of multi-vehicle tests are designed to validate platoon forming and splitting at the intersection, including: vehicles from different platoon numbers splitting after passing the intersection (Multi-vehicles separation test); vehicles from different platoon numbers forming a platoon before passing the intersection (Vehicles not in a group form in a platoon test); and vehicles from the same platoon number forming a platoon at the link (Vehicles in a group form in a platoon test). Parameters such as height $h$, friction $f$ and correlation coefficient $c_{ij}$ can influence the above scenarios. For scenarios considering vehicles from different platoon numbers (Multi-vehicles separation test and Vehicles not in a group form in a platoon test), the platoon forming and splitting are dominated by $p_j^c$, which transforms these scenarios into the single vehicle lateral scenario (Single vehicle lane changing test). As for Vehicles in a group form in a platoon test, vehicles with the same number of platoon or class are required to form a platoon, and $p_{ij}^b$ dominates the platoon formation. Thus, the critical parameter is the correlation coefficient $c_{ij}$. Single vehicle lane changing test, Multi-vehicles separation test, Vehicles not in a group form in a platoon test, and Vehicles in a group form in a platoon test show that the proposed model can generate reasonable behaviors under the reasonable parameter ranges. According to the Single vehicle lane changing test, Multi-vehicles separation test, and Vehicles not in a group form in a platoon test, both height $h$ and friction $f$ can influence the performance of the model, and the reasonable ranges are $h \in[110,150]$ and $f \in[1,4]$. As for Vehicles in a group form in a platoon test, the minimum correlation coefficient $c_{ij}$ in the platoon forming scenario is explored under the reasonable ranges of $h$ and $f$, as shown in Figure~\ref{fig:cali-result}. 
Both the situation of vehicles locating on adjacent lanes and the situation that two vehicles are separated by one lane are tested. The minimum correlation coefficient $c_{ij}$ increases with the growth of height $h$ and friction $f$. Since larger vehicle gaps lead to larger attractions, the minimum correlation coefficient $c_{ij}$ of vehicles in the adjacent lane is larger than the minimum correlation coefficient $c_{ij}$ of vehicles separated by one lane with same values of height $h$ and friction $f$, as shown in Figure~\ref{fig:cali-result}. If vehicles are separated by more than one lane, the minimum correlation coefficient $c_{ij}$ will be much smaller. Suitable value of the minimum correlation coefficient $c_{ij}$ can be selected according to the number of lanes, height $h$ and friction $f$.

 \section{Conclusion} \label{section:conclusion}
This paper presents a flock-like two-dimensional model for cooperative vehicle groups. This model describes  platoon formation process  with  flocking principles under the  mixed traffic environment of CAVs and HVs. The model can be applied to situations where  formation instructions are given by traffic controllers or as self-organizing processes based on vehicle classes. Potential filed is the basis of the model and some rules are added for better adapting to realistic scenarios. The potential field is composed of the inter-vehicle potential filed and the cross-section potential filed. 

The mathematical properties of the model is examined. The car-following behavior is shown to respect the rational car-following behavior. The safety of the proposed model is also analyzed and verified under emergency braking scenario. Seven numerical experiments  are designed to test the performance of the proposed model, and the reasonable ranges of parameters in the flock-like model are explored. In the longitudinal scenarios, the safety of CAV platoon is mainly considered and proved. As for the two-dimensional scenarios, the platoon formation and split behaviors  labeled with different platoon numbers or classes in the vicinity of the intersection are tested, in addition to the platoon formation behavior of vehicles with the same platoon number or class. These scenarios show that the proposed model can mimic the platoon operation in a wide range of maneuvers including normal car-following, emergency braking, platoon merge, and split in the mixed traffic.  

This paper explores platoon operations  under the environment of mixed traffic. Further research will integrate the traffic control algorithm with the proposed platoon dynamics model. The behaviors of human drivers and the trajectory planning of CAVs can be improved in the future. Additionally, the impact of the platoon formation model on traffic oprations can be assessed in terms of capacity and safety.
 
\section*{Acknowledgement(s)}

This research was funded by Shanghai Sailing Program (No. 19YF1451600), the National Natural Science Foundation of China (No. 51722809 and No. 61773293), and the Fok Ying Tong Education Foundation (No. 151076). The views presented in this paper are those of the authors alone.









\bibliographystyle{apalike}
\bibliography{trb_template}

\end{document}